\documentstyle[11pt,epsf]{article}
\input epsf
\begin{document}
\baselineskip 100pt
\renewcommand{\baselinestretch}{01.35}
\renewcommand{\arraystretch}{0.666666666}
{\large
\parskip.2in
\newcommand{\hs}{\hspace{1mm}}
\newcommand{\nhat}{\mbox{\boldmath$\hat n$}}
\newcommand{\cmod}[1]{ \vert #1 \vert ^2 }
\newcommand{\mod}[1]{ \vert #1 \vert }
\newcommand{\pr}{\partial}
\newcommand{\r}{\rho}
\newcommand{\fr}{\frac}
\newcommand{\ie}{{\em ie }}
\newcommand{\th}{\theta}
\newcommand{\p}{\varphi}
\newcommand{\x}{\xi}
\newcommand{\xb}{\bar{\xi}}
\newcommand{\Vd}{V^{\dagger}}
\newcommand{\Ref}[1]{(\ref{#1})}

\title{\hbox{\hspace{18mm}{\Large{\bf Multi-Skyrmion Solutions for the 6th 
order Skyrme Model}}}}

\author{
I.Floratos\thanks{e-mail address: Ioannis.Floratos@durham.ac.uk}
\, and B. Piette\thanks{e-mail address:
 B.M.A.G.Piette@durham.ac.uk}
\\ Department of Mathematical Sciences,University of Durham, \\
 Durham DH1 3LE, UK\\
}\date{March 2001}

\maketitle

\begin{abstract}
Following Marleau \cite{Marl1}, we study an extended version of the 
Skyrme model to which a sixth order term 
has been added to the Lagrangian and we analyse some of its classical 
properties. We compute the multi-Skyrmion solutions numerically for up to 
$B=5$ and show that they have the same symmetries as the usual Skyrmion 
solutions. We use the rational map ansatz introduced by Houghton et al. 
\cite{Manton} to evaluate the energy and the radius for multi-skyrmion 
solutions of up to $B=6$ for both the $SU(2)$ and $SU(3)$ models and
compare these results to the ones obtained numerically.  
We show that the rational map ansatz works as well for the generalised 
model as for the pure Skyrme model.
\end{abstract}

\section{Introduction}

Recent mathematical developments within the area of
non-perturbative methods have established the Skyrme model as the
strongest candidate for an effective low energy theory of quantum
chromodynamics (QCD). The model was originally proposed by T.H.R.
Skyrme \cite{Skyr1} to describe hadron interactions. However, it
was mainly ignored, until it was shown \cite{Hooft,Witt1,Witt2}
that in the large $N_c$ limit, where $N_c$ is the number of
colours, this non-linear theory can describe the low energy limit
of QCD. This revived the Skyrme model and since then significant
progress has been made towards the understanding of its properties
resulting to a relatively successful description of nuclear
interactions.

The Skyrme model is described by an $SU(N)$ valued field $U(\vec{ x},t)$
which must satisfy the boundary condition $U \rightarrow I$ as 
$|\vec{x}| \rightarrow \infty$, where $I$ is the unit matrix. This condition
ensures finiteness of the energy for any field configuration and it also
implies that the three dimensional Euclidean space on which the model is 
defined 
can be compactified into $S^3$. As a result, the Skyrme field $U$ corresponds 
to mappings from $S^3$ into $SU(N)$. Skyrme's idea was to interpret the 
winding number associated with these topologically non trivial mappings
as the baryon charge.

The model is described by the Lagrangian 
\begin{equation} 
\label{lag}
{\cal L}_{Sk}=\frac{F_{\pi}{^2}}{16}
TrR_{\mu}R^{\mu}+\frac{1}{32a^2}Tr[R_{\mu},R^{\nu}][R_{\nu},R^{\mu}]
\end{equation}
where $R_{\mu}=(\partial_{\mu}U)U^{-1}$ is the right chiral current,
$F_{\pi}=189$Mev is the pion decay constant and $a$ is a dimensionless 
parameter. The first term in \Ref{lag} is the non-linear $\sigma$-model and
one can easily show using a scaling argument that with this term alone
static solutions cannot exist. The same argument shows that one must add
to the Lagrangian terms involving higher derivatives. This argument led 
Skyrme to add the second term, usually referred to as the Skyrme term,  
in \Ref{lag} which is the simplest one that preserves the $SU(N)$ and 
Lorentz invariances.

The Skyrme model can be generalised by adding terms involving higher 
order derivatives in the Lagrangian \Ref{lag} 
\cite{Jackson,Marl1,Marl4,Marl5}. 
Doing this, one introduces extra parameters that can be tuned in to
increase the quality of the Skyrme model as an effective low energy limit
of QCD. For example in \cite{Jackson,Adkins2} the sixth-order term was 
used to take into account the $\omega$-meson interactions when computing 
the central Nucleon-Nucleon potential. In a different context Marleau 
studied the model where a large number of higher order terms were included in
the Lagrangian \cite{Marl1,Marl4,Marl5} and where, to avoid the introduction 
of a large number of extra parameters, the coefficients of these extra terms 
were all related to the coefficient of the Skyrme model. 

In this paper we will consider the simplest possible extension of the 
Skyrme model {\it i.e.} defined by the Lagrangian \Ref{lag} to which 
we add the sixth-order term 
\begin{eqnarray}
\label{sixorder}
{\cal L}_6=c_6\,
Tr[R_{\mu},R^{\nu}][R_{\nu},R^{\lambda}][R_{\lambda},R^{\mu}].
\end{eqnarray}
The unknown coefficient $c_6$ denotes the strength of this term
and will be left as a free parameter of the model. This
particular choice of a sixth-order term is not accidental as it is
the only term that preserves the Lorentz invariance and the $SU(N)$ symmetry 
of the model and leads to an equation of motion that does not involve 
derivatives of order higher than two. This is the term that was used in 
\cite{Jackson}.

In this paper we will focus our attention on the static solutions
of the extended Skyrme model and thus  consider fields that do
not depend on time. It is also convenient to define the dimensionless 
parameter $\kappa = 192 c_6 F_\pi^2 a^4$ and to introduce the dimensionless 
units $y = x \sqrt{2}/(a F_\pi)\sqrt{1+\sqrt{1+\kappa}}$ so that the energy 
of the model can be written as
\begin{eqnarray}
\label{energy}
E=-\Lambda\hspace{1mm}\int d\vec{x }\hspace{1mm}^3
  \left(\frac{1}{2} \hspace{1mm}TrR_{i}^2+
        \frac{1-\lambda}{16}\hspace{1mm} Tr[R_{i},R_{j}]^2+
      \frac{1}{96}\lambda\hspace{1mm} Tr[R_{i},R_{j}][R_{j},R_{k}][R_{k},R_{i}]
  \right)
\end{eqnarray}
where $\Lambda=F_{\pi}/(4\sqrt{2} a)\sqrt{1+\sqrt{1+\kappa}}$ and
$\lambda= \kappa /(1+\sqrt{1+\kappa})^2$.
The parameter $\Lambda$ is the energy scale of the model. In what follows it 
will be convenient to use the dimensionless energy expressed in the 
so-called topological units {\it i.e.}
\begin{equation}
\label{energydimless}
\tilde{E} = \frac{E}{12 \pi^2 \Lambda}.
\end{equation} 
We have chosen this parametrisation of the model so that $\lambda \in [0,1]$ 
describes 
the mixing between the Skyrme term and the sixth-order term \Ref{sixorder}. 
When $\lambda=0$ our model reduces to the usual pure Skyrme model while when 
$\lambda = 1 $  the Skyrme term vanishes and the model reduces to what we 
refer to in what follows as the pure Sk6 model. 

The Euler-Lagrange equations derived from \Ref{energy} for the static
solutions are given by
\begin{eqnarray}
\label{eqnR}
\partial_i\left(R_i-\frac{1}{4}(1-\lambda)\Big[R_j,[R_j,R_i]\Big]-
\frac{1}{16}\lambda\Big[R_j,[R_j,R_k][R_k,R_i]\Big]\right)=0.
\end{eqnarray}

As mentioned above, an important property of the Skyrme model is that its
field corresponds to a mapping from $S^3$ into $SU(N)$ and as 
$\pi_3(SU(N)) = Z$ each configuration is characterised by a an integer 
which can be obtained explicitly by evaluating the expression
\begin{eqnarray}
\label{topcharge}
B=\frac{1}{24\pi^2}\int_{R^3}d\vec{x}\hspace{1mm}^3\hs
\varepsilon^{ijk}\hspace{1mm}Tr(R_i \hs R_j \hs R_k ),
\end{eqnarray}
which following Skyrme's idea is interpreted as the baryon number.
Moreover the following inequality holds for every configuration
\begin{eqnarray}
\label{topchargein}
\tilde{E} \ge \sqrt{1-\lambda} B.
\end{eqnarray}

Our extended Skyrme model depends on three parameters: $F_\pi$, $a$ and 
$c_6$ or using the dimensionless units, $\Lambda$, $k$ and $\lambda$.
To determine the physical values for these parameters, we can evaluate
different quantities. As our analysis will be purely classical, we will
use for this purpose the total energy \Ref{energy} and the isoscalar mean 
square matter radius given by \cite{Adkins}
\begin{eqnarray}
R^2\hspace{1mm}\equiv\hspace{1mm}<r^2>_{I=0}=
 \frac{\int^{\infty}_0 dr \hspace{1mm}r^2 \hspace{1mm}\rho_B(r)}
      {\int^{\infty}_0 dr \hspace{1mm}\rho_B(r)}
\end{eqnarray}
where 
\begin{eqnarray}
\rho_B(r)=4\pi r^2\hspace{1mm}B^0(r).
\end{eqnarray}
Notice that after performing the scaling 
$x \rightarrow x \sqrt{2}/(a F_\pi)\sqrt{1+\sqrt{1+\kappa}}$
we can define the matter radius evaluated in dimensionless units as
\begin{equation}
\tilde{R} = \frac{1}{\sqrt{2}\sqrt{1+\sqrt{1+\kappa}}} a F_\pi\, R.
\end{equation}

One can see from the definition of $\tilde{E}$ and $\tilde{R}$ that the
ratio of energy or matter radius for different solutions only depends on 
$\lambda$. In the following section we will evaluate the energy and radius
of multi-Skyrmion solutions for the general model and evaluate these two 
quantities with the corresponding value for the single Skyrmion and
compare them directly to the experimental ratio:    
\begin{eqnarray}
\label{ratio}
\frac{E_{B}}{E_{B=1}}=\frac{\tilde{E}_{B}(\lambda)}{\tilde{E}_{B=1}(\lambda)}
\hspace{5mm} \and \hspace{5mm}
\frac{R_{B}}{R_{B=1}}=\frac{\tilde{R}_{B}(\lambda)}{\tilde{R}_{B=1}(\lambda)}.
\end{eqnarray}

So far all the studies of the classical properties of generalised Skyrme 
models have been focusing  on the properties of the single
skyrmion ($B=1$) \cite{Adkins2,Jackson,Marl1,Marl4,Marl5}. 
In section 2 we compute numerically multi-skyrmion configurations for $B=2$
to $5$ and compare the energy and the radius of these solutions
with the experimental values.

It was shown recently\cite{Manton,sun} that multi-skyrmion configurations, 
{\em ie} $B\geq2$, can be studied systematically using as an approximation 
the so-called harmonic map ansatz.
In section 3 we approximate the multi-Skyrmion solution both for the $SU(2)$ 
and $SU(3)$ model using this ansatz. We compare the results 
obtained with the numerical solutions and we show that the harmonic map 
ansatz provides  a good approximation for the multi-Skyrmion solutions of the 
extended model as well. 

\section{Numerical Solutions}

In this section we investigate the multi-Skyrmion solutions of the extended
$SU(2)$ Skyrme model by solving the static Euler-Lagrange equation 
\Ref{eqnR} of the model numerically. Computing the static 
solutions of such a three-dimensional model is rather difficult and requires
a large amount of computing power. As one has to be very careful when 
assessing the accuracy of such numerical results, we are giving in the 
Appendix a discussion of the numerical methods that we have used.
 
To compute the solution numerically, it is more convenient to describe
the $SU(2)$ fields using a four-component vector $\phi$ of unit length, 
$\cmod{\phi} = 1$, which is related to the unitary field by 
$U \,=\,\phi_0\,I+i\,\vec{\tau}\cdot\vec{\phi}$ where $I$ is
the unit matrix and $\vec{\tau}$ are the Pauli matrices.
The expression for the energy \Ref{energydimless} then becomes
\begin{eqnarray}
\label{energyphi}
\tilde{E}&=&\frac{1}{12\pi^2}\,\int_R  |\phi_{\mu}|^2+
  \frac{1-\lambda}{2}\left[|\phi_{\mu}|^4-(\phi_{\mu}\cdot\phi_{\nu})^2\right] 
\nonumber\\ && 
\hspace{15mm} +\frac{\lambda}{6}\,
   \left[|\phi_{\mu}|^6-3\,|\phi_{\mu}|^2\,(\phi_{\nu}\cdot\phi_{\kappa})^2
      +2\,(\phi_{\mu}\cdot\phi_{\nu})(\phi_{\kappa}\cdot\phi_{\mu})
          (\phi_{\nu}\cdot\phi_{\kappa})\right],
\end{eqnarray}
and the Euler-Lagrange equations derived from \Ref{equationphi}, after adding
a Lagrange multiplier to impose the constraint $\cmod{\phi} = 1$, are given 
by 
\begin{eqnarray}
\label{equationphi}
& & \phi_{\mu\mu}\,
\Bigg(1+(1-\lambda)\,|\phi_{\nu}|^2
      +\frac{1}{2}\lambda\,|\phi_{\nu}|^2|\phi_{\kappa}|^2
      -\frac{1}{2}\lambda\,(\phi_{\nu}\cdot\phi_{\kappa})^2
\Bigg)
+|\phi_{\mu}|^2\cdot\phi
\nonumber \\
&& +(1-\lambda) 
   \left((\phi_{\nu}\cdot\phi_{\nu\mu})\,\phi_{\mu}
         -(\phi_{\mu\mu}\cdot\phi_{\nu})\,\phi_{\nu}
         -(\phi_{\mu}\cdot\phi_{\nu})\,\phi_{\nu\mu}
         +|\phi_{\mu}|^4\,\phi-(\phi_{\mu}\cdot\phi_{\nu})^2\phi
    \right)
\nonumber \\ 
&& +\lambda\,\Bigg(
 \phi_{\mu}\,(\phi_{\nu}\cdot\phi_{\nu\mu})\,|\phi_{\kappa}|^2-
 \phi_{\mu}\,(\phi_{\nu}\cdot\phi_{\kappa})\,(\phi_{\nu}\cdot\phi_{\kappa\mu})-
 \phi_{\nu}\,(\phi_{\mu}\cdot\phi_{\mu\kappa})\,(\phi_{\nu}\cdot\phi_{\kappa})
\nonumber\\ 
&& \hspace{12mm}
 -\phi_{\nu}\,(\phi_{\nu}\cdot\phi_{\kappa\kappa})\,|\phi_{\mu}|^2
 -\phi_{\nu\kappa}\,(\phi_{\nu}\cdot\phi_{\kappa})\, |\phi_{\mu}|^2
 +\phi_{\nu\mu}\,(\phi_{\kappa}\cdot\phi_{\mu})\,(\phi_{\nu}\cdot\phi_{\kappa})
\nonumber \\ 
&& \hspace{12mm}
+\phi_{\nu}\,(\phi_{\kappa}\cdot\phi_{\mu\mu})\,(\phi_{\nu}\cdot\phi_{\kappa})+
 \phi_{\nu}\,(\phi_{\kappa}\cdot\phi_{\mu})\,(\phi_{\nu}\cdot\phi_{\kappa\mu})
\nonumber \\ 
&& \hspace{8mm}+
 \frac{1}{2}\,\left[|\phi_{\mu}|^6
  -3\,|\phi_{\mu}|^2\,(\phi_{\nu}\cdot\phi_{\kappa})^2
  +2\,(\phi_{\mu}\cdot\phi_{\nu})(\phi_{\kappa}\cdot\phi_{\mu})
      (\phi_{\nu}\cdot\phi_{\kappa})\right]\,\phi \,\,
\Bigg)=0.
\end{eqnarray}

To compute the $B=1$ solution, we use the so-called hedgehog ansatz 
\begin{eqnarray}
\label{hedgehog}
\phi=\left(
     \begin{array}{l}
       \sin f(r) \hspace{1mm}\sin \theta\hspace{1mm}\sin(\varphi)\\ 
       \sin f(r) \hspace{1mm}\sin \theta\hspace{1mm}\cos(\varphi)\\ 
       \sin f(r) \hspace{1mm}\cos \theta \\ 
       \cos f(r)
     \end{array} \right)
\end{eqnarray}
where $r$, $\theta$ and $\varphi$ are the usual spherical coordinates.
Plugging \Ref{hedgehog} into \Ref{energyphi} one minimises the energy for
the profile function $f(r)$ which then has to satisfy an ordinary differential
equation. This is a very special case of the harmonic map ansatz discussed in 
the next section, so we will just say at this stage that the solutions are 
radially symmetric and that the $\lambda$ dependence of the energy and the 
radius of the solutions are given on Figure 1. The fact that the energy 
decreases with $\lambda$ is entirely due to our choice of parametrisation;
the real quantities one has to look at are the energy and radius ratio 
\Ref{ratio}.    

\vskip 5mm
\begin{figure}[htbp]
\unitlength1cm \hfil
\begin{picture}(16,6)
 \epsfxsize=8cm \epsffile{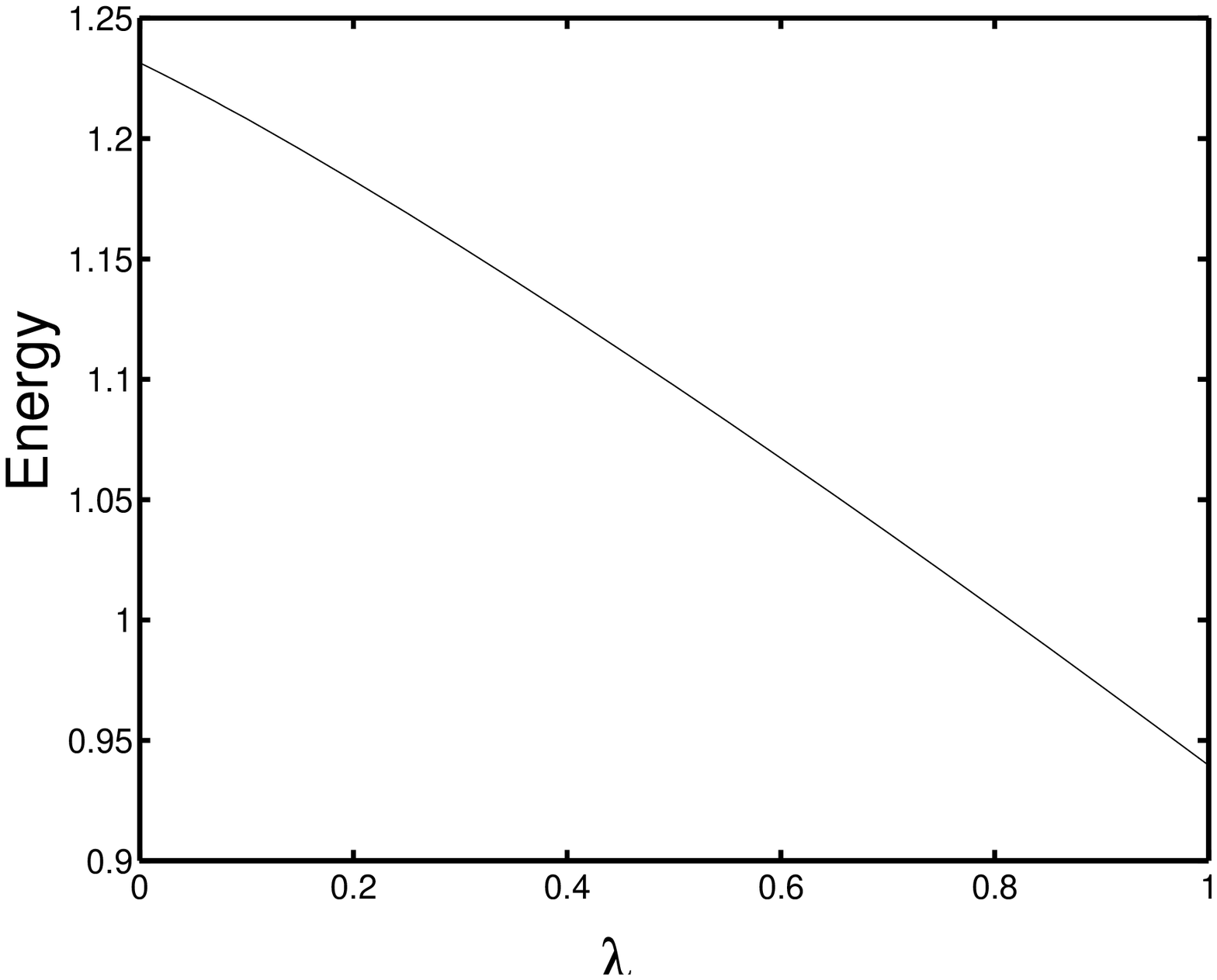}
 \epsfxsize=8cm \epsffile{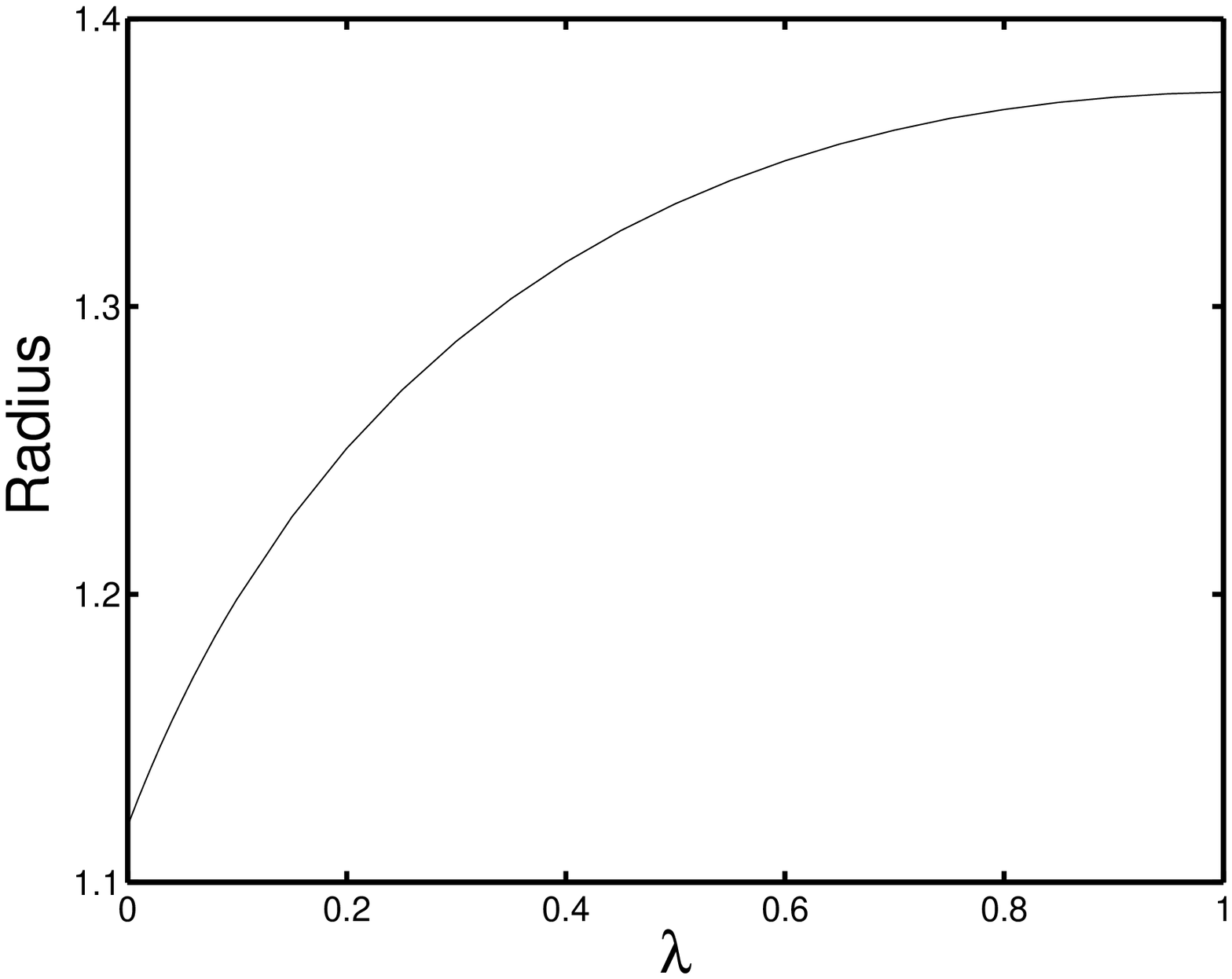}
\end{picture}
\caption{ $\tilde{E}(\lambda)$ and $\tilde{R}(\lambda)$ for the 
$B=1$ solutions.}
\end{figure}

As described in the Appendix, solving \Ref{equationphi} accurately is rather 
difficult. For this reason the case $B=2$ was solved differently. It is indeed
well known that the usual $B=2$ static solution is axially symmetric 
\cite{BS,axi3,axi1,axi2} and we found that this is also true for the extended
Skyrme model. Knowing this, we can reduce the system of equations for these
solutions to a two-dimensional system by using the ansatz
\begin{eqnarray}
\label{phi2}
\phi=\left(
     \begin{array}{l}
       \sin f \hspace{1mm}\sin g\hspace{1mm}\sin(2\varphi)\\ 
       \sin f \hspace{1mm}\sin g\hspace{1mm}\cos(2\varphi)\\ 
       \sin f \hspace{1mm}\cos g \\ 
       \cos f
     \end{array} \right)
\end{eqnarray}
where $\varphi = \mbox{atan}(y/x)$. The two profile functions,  
$f(\rho,z)$ and $g(\rho,z)$ are functions of the usual axial coordinates 
$\rho = \sqrt{x^2+y^2}$ and $z$ and they satisfy the following boundary 
conditions:
\begin{eqnarray}
\begin{array}{rlrlrl}
f(0,0) &=\,\pi & \qquad f(\rho\rightarrow\infty,z\rightarrow\infty) &=\, 0& 
\qquad f_\rho(0,z) &=\,0\\
g(0,z < 0) &=\,0 & \qquad  g(0,z > 0) &=\, \pi& 
\qquad  g_R\vert_{R\rightarrow\infty} &=\,0
\end{array}
\end{eqnarray}
where $R^2 = r^2 + z^2$. 

Substituting \Ref{phi2} into \Ref{energyphi} we get
\begin{eqnarray}
\label{energyax}
E &=& \frac{1}{6\pi}\,\,\,\int
         \Bigg\{\bigg[(f_{\rho}^2+f_{z}^2)+\sin^2f\,(g_{\rho}^2
                      +g_{z}^2)+\frac{4}{\rho^2}\,\sin^2f\,\sin^2g \bigg]  
\nonumber \\ 
&&+(1-\lambda)\bigg[\frac{4}{\rho^2}\,\sin^2fsin^2g\big[f_{\rho}^2
         +f_z^2+\sin^2f(g_{\rho}^2+g_z^2)\big]
         +\sin^2f\left(f_{\rho}g_z-f_zg_{\rho}\right)^2
   \bigg]
\nonumber \\
 & & \hspace{25mm}+\lambda\hspace{1mm}
 \bigg[ \frac{4}{\rho^2}\,\sin^4f\hspace{1mm}\sin^2g
        \left(f_{\rho}g_z-f_z g_{\rho}\right)^2 
 \bigg]
\Bigg\}\hspace{2mm}\rho \hspace{2mm}d\rho dz
\end{eqnarray}
and the corresponding Euler-Lagrange equations are given by 
\begin{eqnarray}
\label{equationaf}
&& \left(f_{\r\r}+f_{zz}+\fr{1}{\r} f_\r\right)
    -\fr{2}{\r^2}\sin2f\,\sin^2g-\fr{1}{2}\sin 2f\,(g^{2}_\r+g^{2}_z)
\nonumber \\ 
&& +(1-\lambda) 
     \Bigg\{\fr{4}{\r^2}\sin^2f\,\sin2g\,\left(f_\r g_\r+f_z g_z \right)+ 
 \fr{1}{\r}\,\sin^2f\,\left(f_\r g_{z}^2-f_z g_\r g_z\right)
 +\fr{1}{2}\sin 2f\,\left( f_\r g_z-f_z g_\r\right)^2
\nonumber \\ 
&&   +\fr{4}{\r^2}\,\sin^2f\,\sin^2g
   \left(f_{\r\r}+f_{zz}-\fr{1}{\r}f_\r\right)+ 
   \fr{4}{\r^2}\sin2f\,\sin^2g\,
   \left[\fr{1}{2}(f^{2}_\r+f^{2}_z)-\sin^2f\,(g^{2}_\r+g^{2}_z )\right] 
\nonumber \\ 
&& + \sin^2f\,\Big\{f_{\r\r}g_{z}^2+f_{zz}g^{2}_\r-2f_{z\r}g_zg_\r+ 
    f_\r g_z g_{z\r}-f_z g_{\r\r}  g_z-f_\r g_{zz} g_\r+f_z g_\r g_{\r z}
     \Big\}
     \Bigg\} + 
\nonumber \\ 
&& \lambda\,\Bigg\{\fr{8}{\r^2}\sin^2f\, \sin 2f\, \sin^2g
    \left(f_\r g_z-f_z g_\r\right)^2+\fr{4}{\r^2}\sin^4f\,\sin^2g
    \Big(f_{\r\r}g^{2}_z+f_\r g_z g_{z\r}- 
\nonumber \\ 
&& -2f_{z\r}g_\r g_z-f_z g_{\r\r}g_z+f_{zz}g^{2}_\r+
    f_z g_\r g_{\r z}-f_\r g_{zz}g_\r-
    \fr{1}{\r}(f_\r g^{2}_z-f_z g_\r g_z )\Big)- 
\nonumber \\ 
&& \hspace{60mm} \fr{4}{\r^2}\sin2f \, \sin^2f \, \sin^2g\, 
    \left(f_\r g_z-f_z g_\r \right)^2 \Bigg\}=0
\end{eqnarray}
and
\begin{eqnarray}
\label{equationag}
    \left(g_{\r\r}+g_{zz}+\fr{1}{\r} g_\r\right)+
    \fr{\sin2f}{\sin^2 f}\left(f_\r g_\r+f_z g_z\right)+\fr{2}{\r^2}\sin2g
    +(1-\lambda) \Bigg\{
       \fr{4}{\r^2}\sin^2f\,\sin2g\,\left(g^{2}_\r+g^{2}_z\right)+
&&\nonumber \\
  \fr{4}{\r^2}\,\sin^2g\,
   \Big( 2\sin2f\,\left(f_\r g_\r+f_z g_z\right)+
         \sin^2f \left(g_{\r\r}+g_{zz}-\fr{g_\r}{\r}\right)
   \Big) 
   +\fr{1}{\r}\left(f^{2}_z g_{\r} -f_\r f_z g_{z} \right) 
   +f^{2}_z g_{\r\r}+ f^{2}_\r g_{zz}
&&\nonumber\\ 
    -\fr{2}{\r^2}\sin2g
     \Big\{f^{2}_\r+f^{2}_z+\sin^2f\left(g^{2}_\r+g^{2}_z\right) \Big\} 
    +f_z f_{z \r} g_{\r}-f_{\r\r} g_z f_{z}+
    f_\r f_{\r z} g_{z}-2f_\r g_{z\r} f_{z}-f_{zz}g_\r f_\r\Bigg\}
&&\nonumber \\ 
    +\lambda\, \Bigg\{\fr{4}{\r^2}\sin^2f\,\sin2g\left(f_z g_\r -f_\r g_z
                                                 \right)^2 
    +\fr{4}{\r^2}\sin^2f \, \sin^2g
    \Big[f_z f_{z \r}g_\r+ f_{z}^2 g_{\r\r}-f_{\r\r} f_{z}g_z
    -2f_\r f_{z}g_{z \r}+
&&\nonumber \\ 
    f_\r f_{\r z}g_z +f^{2}_\r g_{zz}-f_{zz} f_\r g_\r
   -\fr{1}{\r}\left(f^{2}_z g_\r-f_\r f_z g_z \right)\Big]-
     \fr{2}{\r^2}\sin^2f\,\sin2g \left(f_\r g_z - f_z g_\r \right)^2 
     \Bigg\} =0.\qquad
&&
\end{eqnarray}
The advantage of having a two-dimensional system is that we can use much 
larger grids and obtain much more accurate results. As discussed in the 
Appendix, we have also compared the $B=2$ solutions obtained by solving
\Ref{equationphi} and \Ref{equationaf},\Ref{equationag} in order to evaluate 
the accuracy of the method we used to solve \Ref{equationphi} numerically. 

In figures 2 to 5, we present the $\lambda$ dependence of the energy and 
radius ratio for the $B = 2$ to $B = 5$ multi-Skyrmion solutions. 
We see that in each case the energy ratio decreases when the coefficient of 
the sixth-order term increases while on the other hand, the radius ratio 
increases  thus making the multi-Skyrmion solution broader in all cases except 
for $B=2$.  
Tables 1 and 2 compare the energy and radius ratio of the pure Skyrme and 
the pure Sk6 models with the experimental values.
We notice that the predicted values for the energy are smaller than the 
experimental values and that the addition of the sixth-order term 
makes the energy ratio even smaller. On the other hand, the addition of 
the  sixth-order term makes the multi-Skyrmion solution broader, except when
$B=2$, but the actual values are still much smaller than the experimental ones.

Another observation we made is that the symmetries of the multi-Skyrmion
solutions for the general model are the same as for the pure Skyrme model.
The solutions for $B=2,3$ and $4$ Skyrmion have respectively the shape of a 
torus, a tetrahedron and a cube while the $B=5$ Skyrmion solution 
has the same $D_{2d}$ symmetry. 

It is a well know problem that the binding energies 
predicted by the Skyrme model are too large and that the radius of the 
classical solutions are too small. One usually argues that quantising the 
model will 
somewhat solve this problem. Adding the sixth-order term does not improve
this: the energy binding is even stronger and the multi-Skyrmion solutions
are narrower except for B=2.

\vskip 5mm
\begin{figure}[htbp]
\unitlength1cm \hfil
\begin{picture}(16,6)
 \epsfxsize=8cm \epsffile{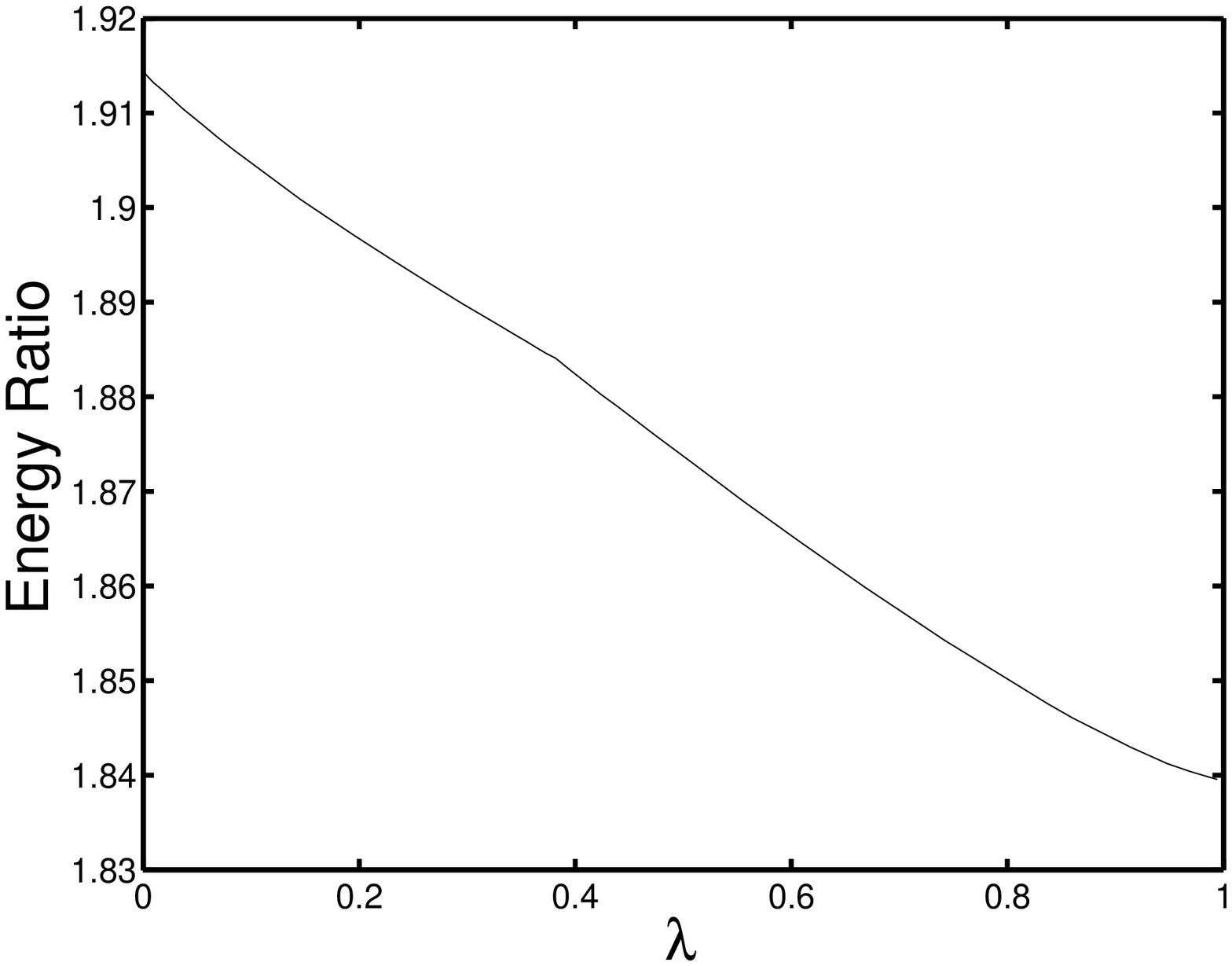}
 \epsfxsize=8cm \epsffile{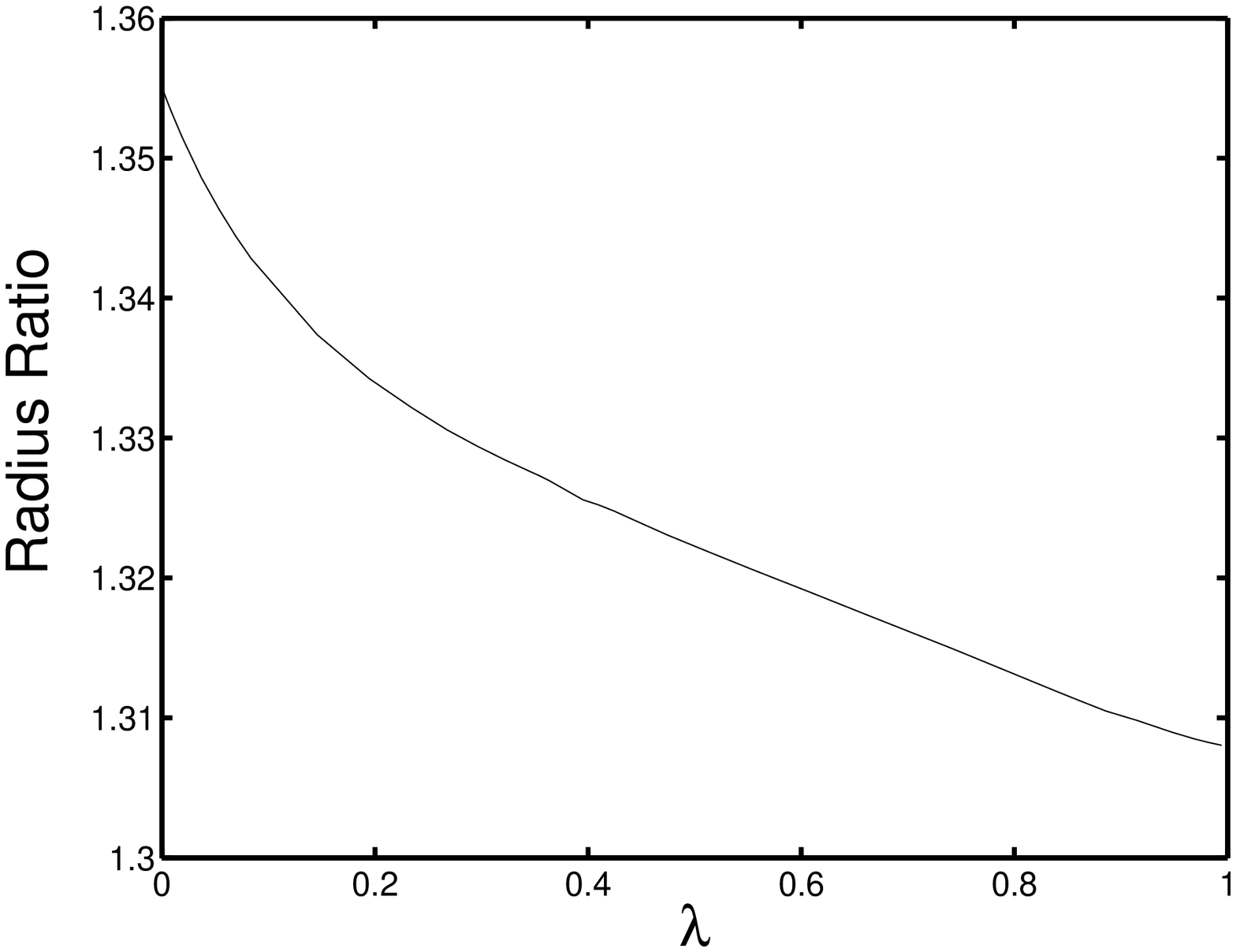}
\end{picture}
\caption{ $\tilde{E}(\lambda)$ and $\tilde{R}(\lambda)$ ratio of
$\hspace{2mm}B=2/B=1\hspace{1mm}$ for the numerical solutions.}
\end{figure}

\vskip 5mm
\begin{figure}[htb]
\unitlength1cm \hfil
\begin{picture}(16,6)
 \epsfxsize=8cm \epsffile{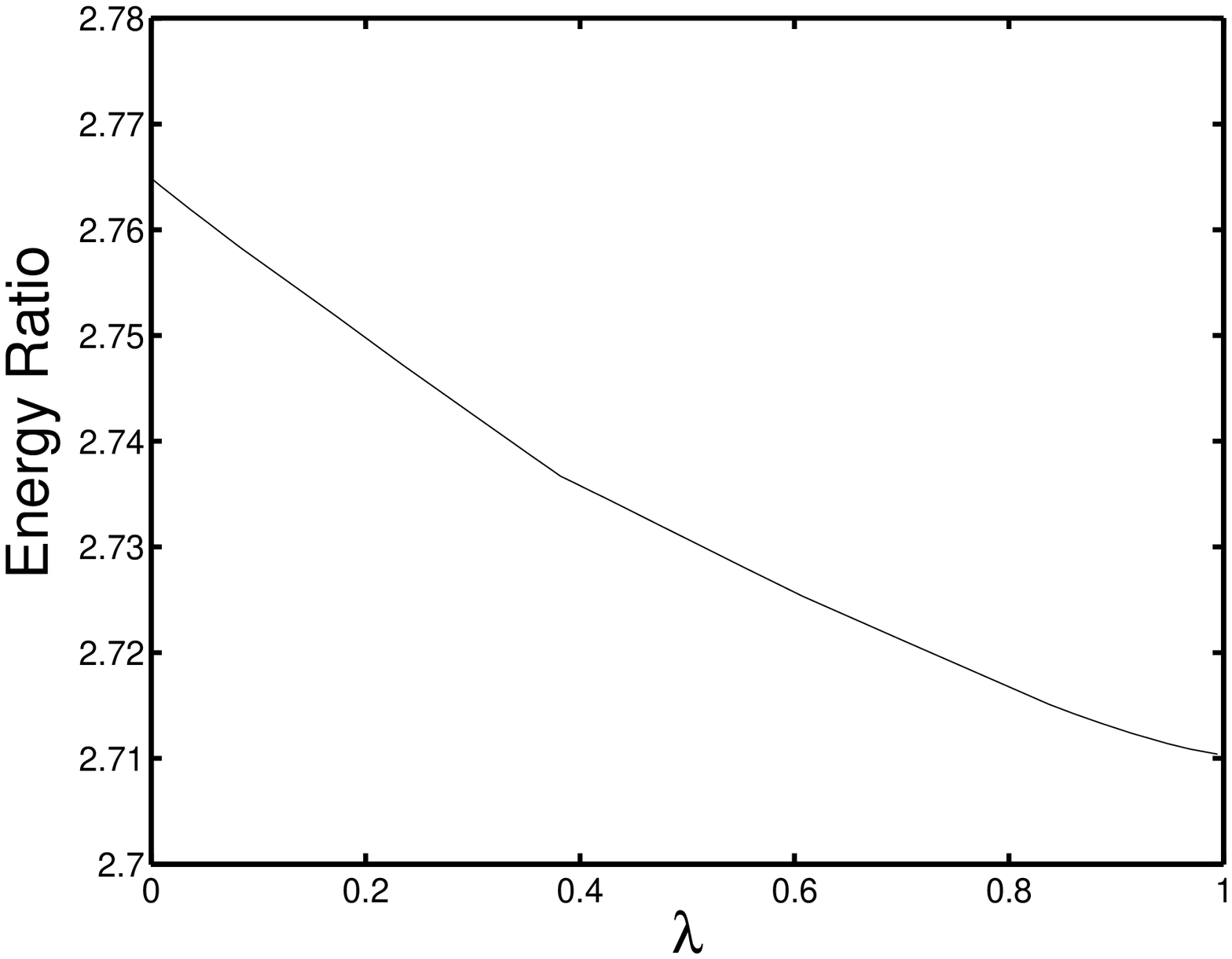}
 \epsfxsize=8cm \epsffile{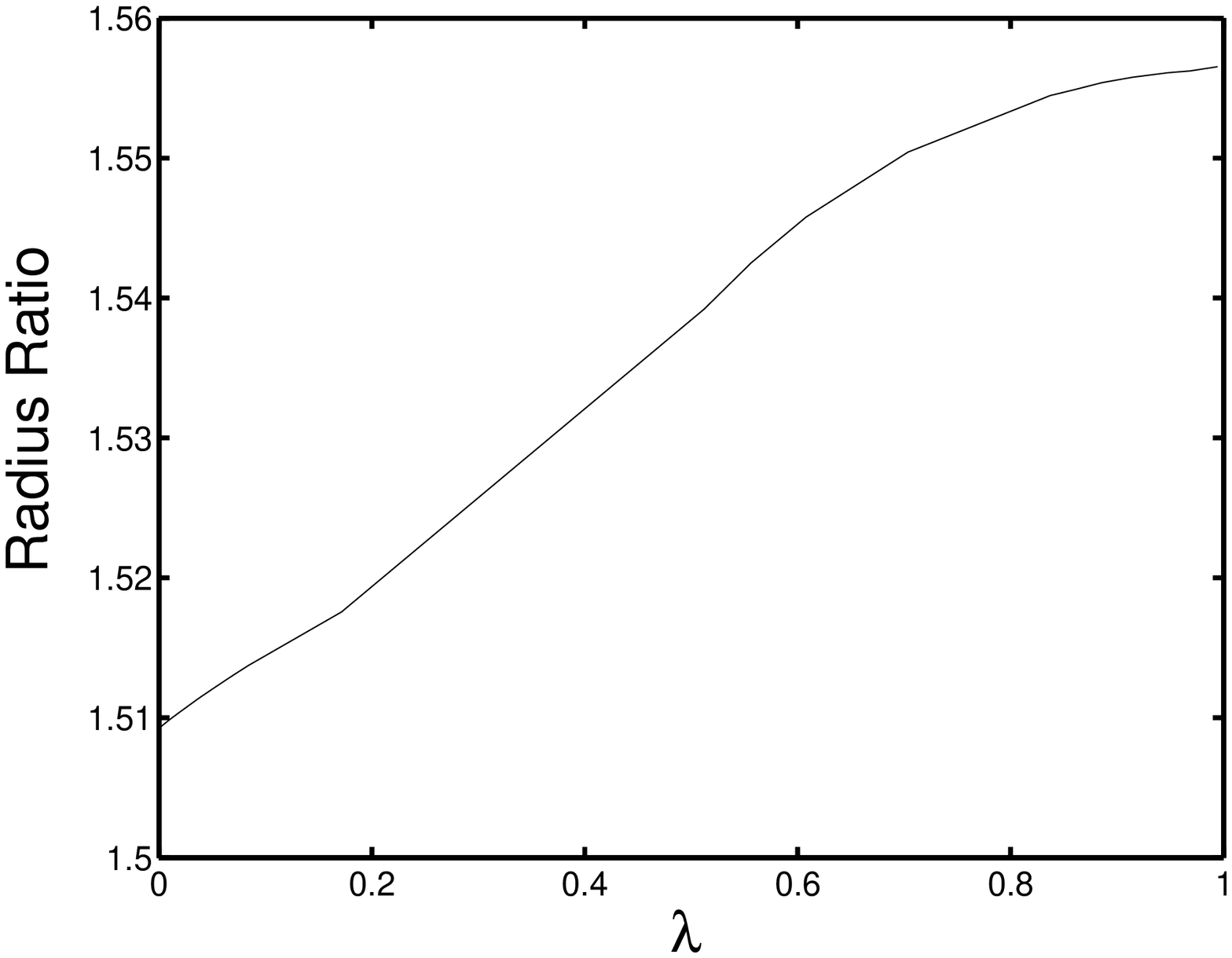}
\end{picture}
\caption{ $\tilde{E}(\lambda)$ and $\tilde{R}(\lambda)$ ratio of
$\hspace{2mm}B=3/B=1\hspace{1mm}$ for the numerical solutions.}
\end{figure}

\vskip 5mm
\begin{figure}[htb]
\unitlength1cm \hfil
\begin{picture}(16,6)
 \epsfxsize=8cm \epsffile{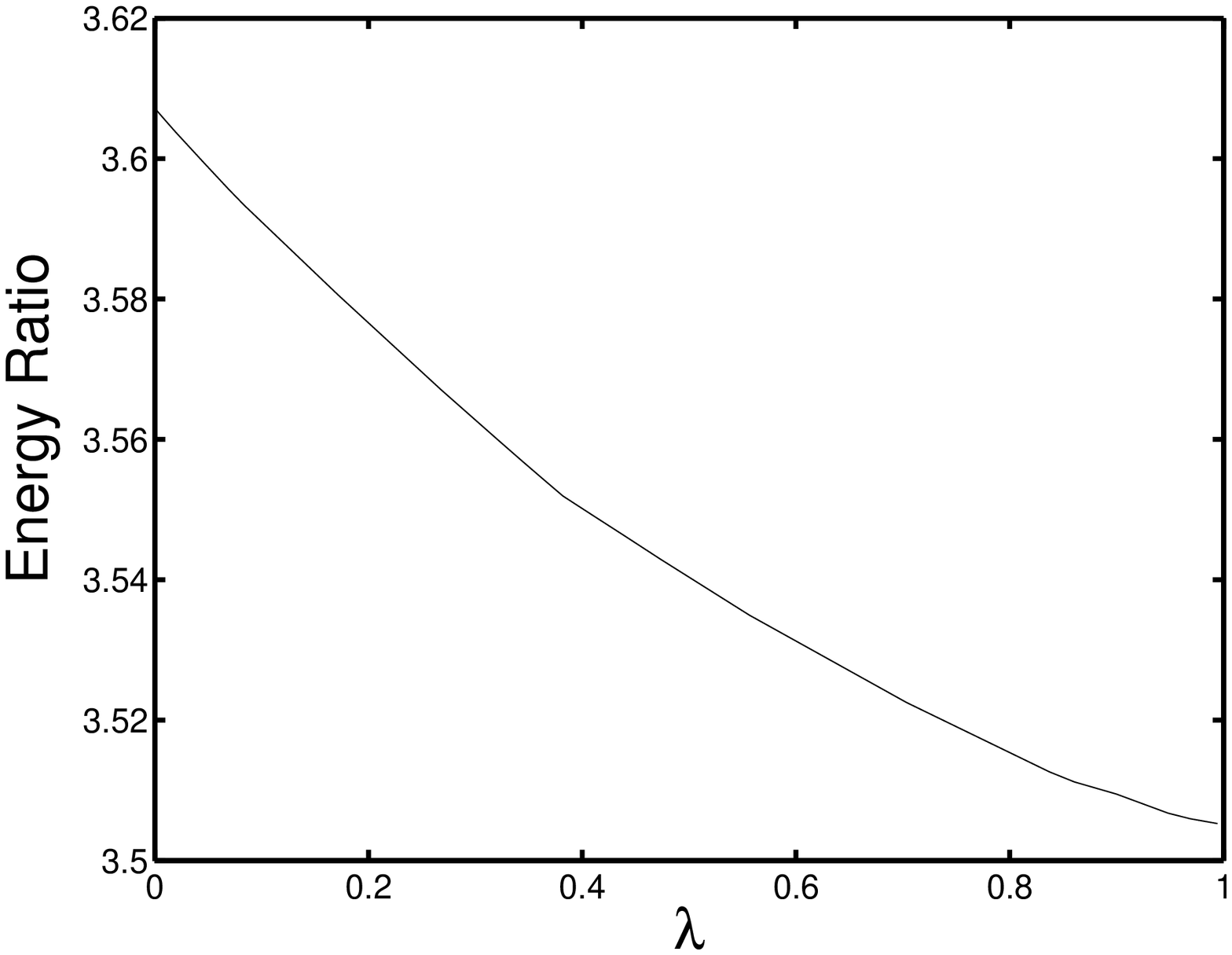}
 \epsfxsize=8cm \epsffile{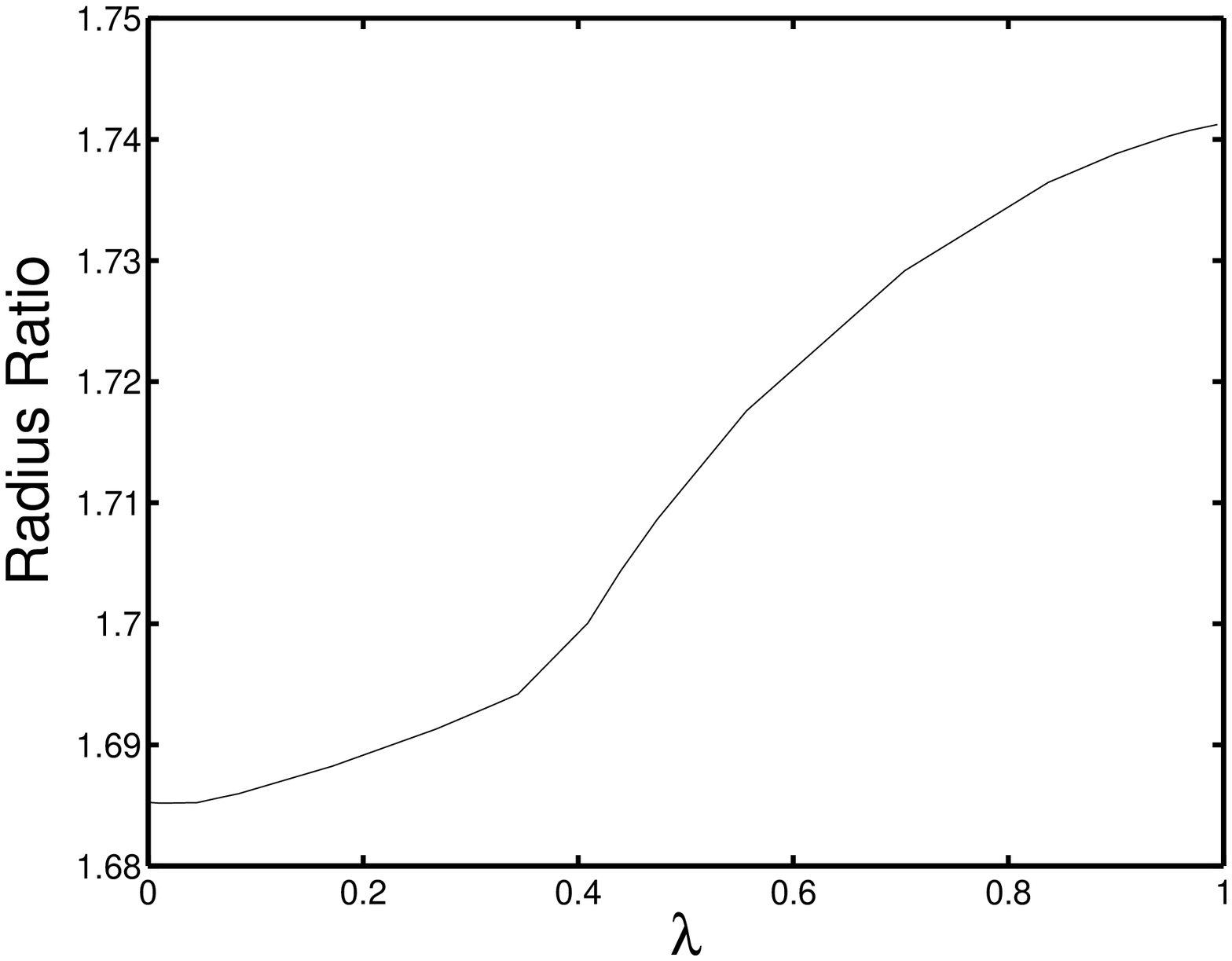}
\end{picture}
\caption{ $\tilde{E}(\lambda)$ and $\tilde{R}(\lambda)$ ratio of
$\hspace{2mm}B=4/B=1\hspace{1mm}$ for the numerical solutions.}
\end{figure}

\vskip 5mm
\begin{figure}[htbp]
\unitlength1cm \hfil
\begin{picture}(16,6)
 \epsfxsize=8cm \epsffile{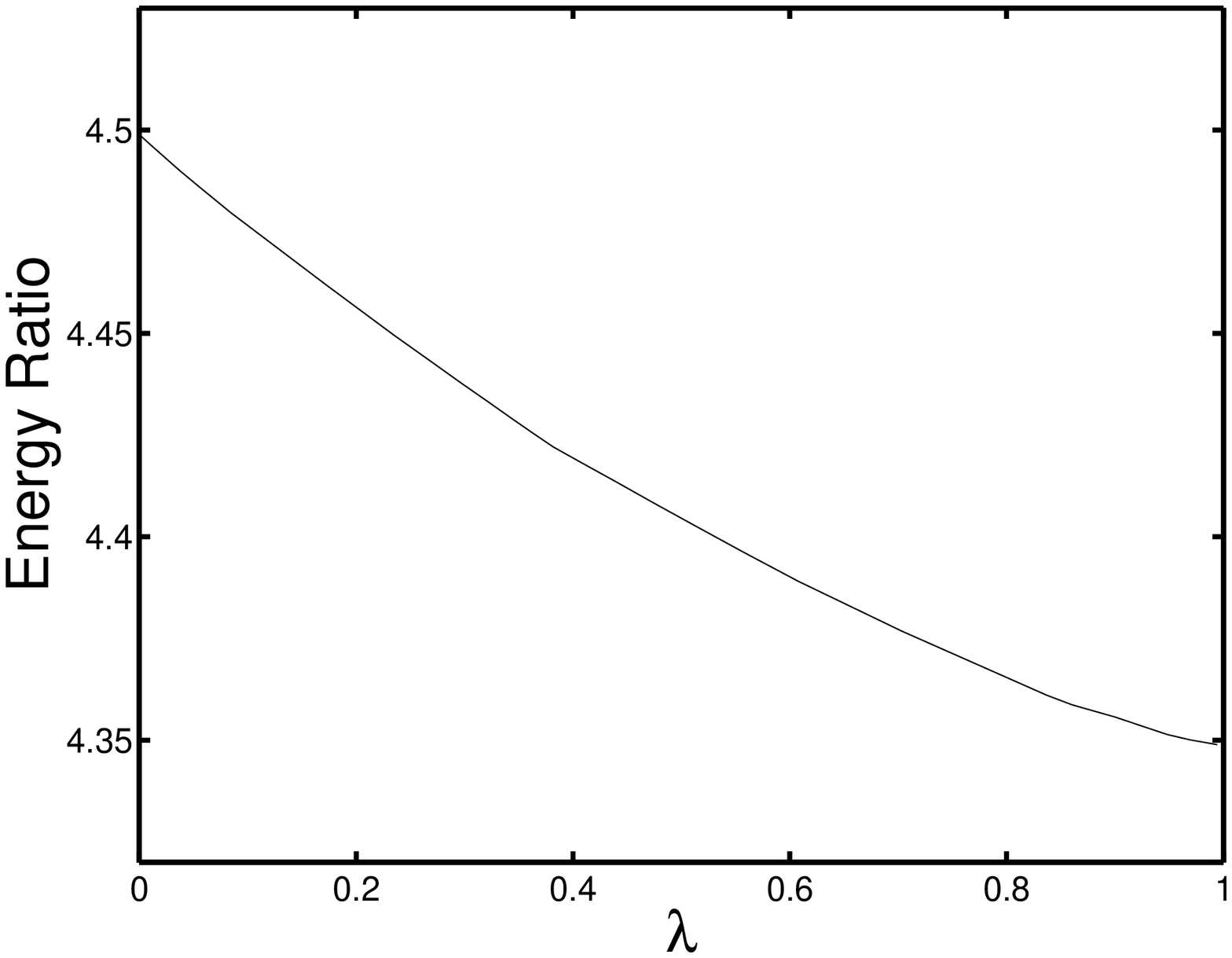}
 \epsfxsize=8cm \epsffile{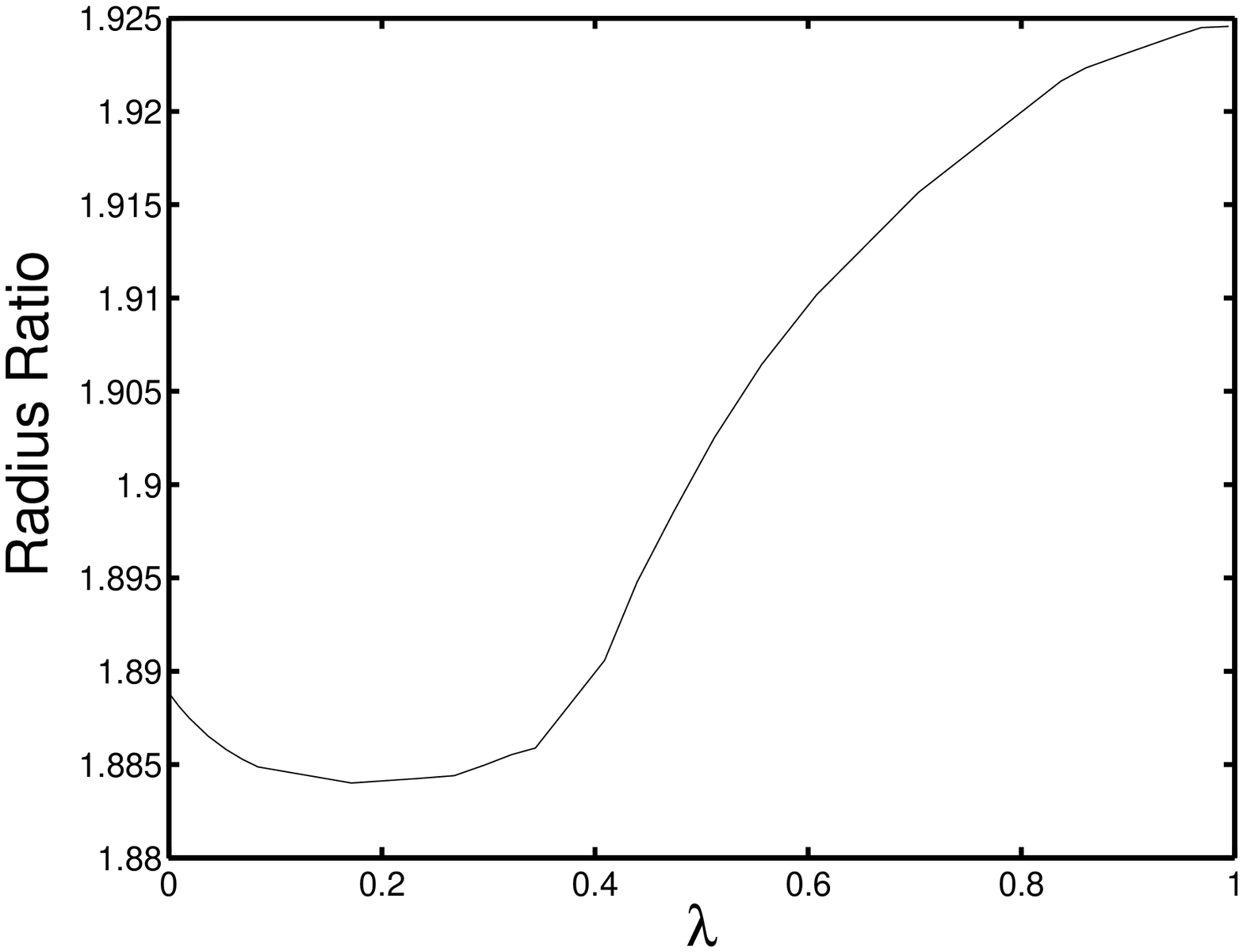}
\end{picture}
\caption{ $\tilde{E}(\lambda)$ and $\tilde{R}(\lambda)$ ratio of
$\hspace{2mm}B=5/B=1\hspace{1mm}$ for the numerical solutions.}
\end{figure}

\begin{table}[ht]
\begin{center}
\begin{tabular}{|c||c |c||c|c| }
 \hline
 & \multicolumn{2}{c||}{} &  \multicolumn{2}{c|}{}
\\& \multicolumn{2}{c||}{Experiment} & \multicolumn{2}{c|}{Numerical solutions}
\\ \hline & & & &  \\ $B$ & Energy (MeV) & Ratio & Skyrme Ratio &
Sk6 Ratio
\\\hline & & & &  \\

1 & 939      & - & - & - \\

2 & 1876.1   & 1.99798 & 1.9009&1.8395\\

3 & 2809.374 & 2.99188 & 2.7650& 2.7103\\

4 & 3728.35  & 3.97055 & 3.6090 & 3.5045\\

5 & 4668.795 & 4.97209 & 4.5000& 4.3780\\

\hline
\end{tabular}
\end{center}
\caption{Experimental energy ratio, $E_B/E_{B=1}$, and values obtained for 
the numerical solutions. The experimental values (MeV)  correspond to 
isotopes with minimum mass \cite{Cohen}.}
\end{table}

\begin{table}[ht]
\begin{center}
\begin{tabular}{|c||c |c||c|c| }
 \hline
 & \multicolumn{2}{c||}{} &  \multicolumn{2}{c|}{}
\\& \multicolumn{2}{c||}{Experiment} & \multicolumn{2}{c|}{Numerical solutions}
\\ \hline & & & &  \\ $B$ & Radius (fm) & Ratio & Skyrme Ratio &
Sk6 Ratio
\\\hline & & & &  \\
1 & 0.72      & - & - & - \\

2 & 1.9715& 2.73819  &1.3549 & 1.308 \\

3 & 1.59& 2.2083  & 1.5080& 1.5570 \\

4 & 1.49& 2.06944 &1.6850 &1.7420  \\

5 & - & - &  1.8890& 1.9250 \\ \hline
\end{tabular}
\end{center}
\caption{Experimental radius ratio, $R_B/R_{B=1}$ and values obtained for 
the numerical solutions. The experimental values (fm)  correspond to nuclei 
with minimum mass. \cite{Adkins,Martorell,Tanihata1,Eg}}
\end{table}

\section{Harmonic map ansatz}

In this section we will use the rational map ansatz to compute configurations
that approximate solutions of the extended Skyrme model. We will then use
these configurations to evaluate the energy and radius of the multi-Skyrmion
configurations, check how these properties depend on $\lambda$ and compare 
these results to the ones obtained for the numerical solutions.

The rational map ansatz, introduced by Houghton et al. \cite{Manton} is an 
extension of the hedgehog ansatz found by Skyrme, which using the usual 
polar coordinates is given by 
\begin{equation}
U(r,\theta,\varphi) = \exp(i g(r)\, \mbox{\boldmath 
$\hat n(\theta,\varphi) \cdot \sigma$}).
\end{equation}
In the hedgehog ansatz $\hat n$ is a unit length vector describing the 
one-to-one mapping of the two-sphere into itself and $g(r)$ is a profile 
function satisfying the boundary conditions $g(0) = \pi$ and $g(\infty) = 0$.
The rational map ansatz consists in using for $\hat n(\theta,\varphi)$ 
harmonic maps from 
$S^2$ into $S^2$ while keeping the same boundary conditions for $g$. One can 
easily show that the baryon number for such a configuration is given by the 
degree of the harmonic map. To approximate a solution of a given baryon 
charge, one takes for $\hat n$ the most general rational map of the given 
degree, inserts the ansatz into the expression for the energy and tries to 
minimise this expression with respect to the parameters of the rational map 
and the profile functions $g$. 
When doing so, the integration over the radius $r$ and the angular 
variables $\theta$ and $\varphi$ decouple and the rational map appears only in
two expressions integrated over the whole sphere. One of them can be evaluated 
explicitly and is equal to the degree of the harmonic map while the other
must be minimised with respect to the parameters of rational map. Doing so
leads to a unique rational map, up to an arbitrary rotation, which describes 
the radial dependence of the Skyrmion configuration.  Then one minimises the 
effective energy by solving the Euler-Lagrange equation for the profile 
function $g$. The configurations obtained by this construction have the 
same symmetries as the exact solutions \cite{BS} and their energies are 
only 1 or 2 percent higher \cite{Manton}.

This construction was later generalised by Ioannidou et al. \cite{sun}
to approximate solutions of the $SU(N)$ Skyrme model using harmonic maps
from $S^2$ into $CP^{N-1}$. The generalised ansatz takes the form   
\begin{eqnarray}
\label{genansatz}
U(r,\theta, \varphi)&=&e^{2ig(r)(P(\theta, \varphi)-{\it I}/N)}\nonumber\\
        &=&e^{-2ig(r)/N}\left({\it I}+(e^{2ig(r)}-1)P(\theta, \varphi)\right)
\end{eqnarray}
where $P(\theta, \varphi)$ is an $N\times N$ projector. As we want to study
some solutions of the $SU(3)$ model as well, we will use the generalised 
construction.
At this stage it is convenient to introduce the complex coordinate
$\xi=\tan(\theta/2)e^{i\varphi}$ which corresponds to the stereographic
projection of the unit sphere onto the complex plane.

The procedure to minimise the energy is the same as the one outlined above 
where the projector will be taken as a harmonic map from $S^2$ into 
$CP^{N-1}$ {\it i.e.} a projector of the form \cite{Din} 
\begin{eqnarray}
\label{projector}
P(f)=\frac{f \otimes f^\dagger}{|f|^2}
\end{eqnarray}
where $f$ is a $N$ components complex vector whose entries are all 
rational functions of $\xi$. The degree of the harmonic map is given by the
highest degree of the components of $f$ and the baryon number is again
given by that degree.

Substituting the ansatz \Ref{genansatz} in the general expression of the
energy density \Ref{energydimless} we get
\begin{eqnarray}
\label{energyans}
\tilde{E}={1 \over 3 \pi} \int dr \hspace{3mm} 
 (A_N\, g_r^2 \,r^2 + 2{\cal N}\, \sin^2g\, (1+(1-\lambda)g_r^2) 
       +(1-\lambda) {\cal I}\, \frac{\sin^4g}{r^2} && \nonumber\\ 
\hspace{30mm} +\lambda\,{\cal I}\,\frac{\sin^4g}{r^2}\,g_r^2
               +\frac{2}{3} \, \lambda\,{\cal M}\,\frac{\sin^6g}{r^4})
\end{eqnarray}
where :
\begin{eqnarray}
\label{minpars}
 A_N&=& \frac{2}{N}(N-1),\nonumber\\
 {\cal N}&=&\frac{i}{2\pi}\int d\xi\,d\bar{\xi}\,\mbox{Tr}
            \left(|\pr_\xi P|^2\right),\nonumber\\ 
 {\cal I}&=&\frac{i}{4\pi}\int d\xi \,d\bar{\xi}\, (1+|\xi|^2)^2\,\mbox{Tr}
            \left([\pr_\xi P,\,\pr_{\bar{\xi}} P]\sp2\right),\nonumber\\
 {\cal M}&=&\frac{i}{8\pi}\int d\xi \,d\bar{\xi}\, (1+|\xi|^2)^4\,\mbox{Tr}
            \left([\pr_\xi P,\,\pr_{\bar{\xi}} P]\sp3\right).
\end{eqnarray}
The integral $\cal N$ is nothing but the energy of the two-dimensional 
Euclidean $CP^{N-1}$ \mbox{$\sigma$-model} and for the harmonic projector it 
is equal to the degree of the harmonic map, 
\begin{eqnarray}
B=\frac{i}{2\pi}\int d\xi\,d\bar{\xi}\,\mbox{Tr}
   \left(P\,[\pr_{\bar{\xi}} P,\,\pr_{\xi}P]\right).
\end{eqnarray}
As $\cal I$ and $\cal M$ 
are independent of $r$ they can be minimised with respect to the
parameters of the harmonic maps. In what follows we will prove that
${\cal M}$ is identically zero so only $\cal I$ will have to be minimised,
something which was already done in \cite{Manton} and \cite{sun}.
The minimisation of the energy with respect to the profile function $g(r)$  
is then  straightforward.

To prove that ${\cal M}$ vanishes, we need to use some properties of the
projectors given by \Ref{projector} where 
$\partial f / \partial \bar{\xi} = 0$. First of all, it is easy to check
that 
\begin{eqnarray}
PP_\xi=0 \hspace{10mm} \mbox{and} \hspace{10mm} P_\xi P=P_{\xi}
\end{eqnarray}
where $P_\xi$ denotes the derivative of $P$ with respect to $\xi$ and
from this we have
\begin{eqnarray}
\label{pprop}
PP_{\xi}P=0 \hspace{5mm}\mbox{and thus} \hspace{5mm} P_{\xi}^2=0.
\end{eqnarray}
Using \Ref{pprop} we notice that 
\begin{eqnarray}
Tr[P_{\xi},P_{\bar{\xi}}]^n
\,&=&\,Tr(P_{\xi}P_{\bar{\xi}}-P_{\bar{\xi}}P_{\xi})^n\nonumber\\
\,&=&\,Tr\left((P_{\xi}P_{\bar{\xi}})^n+(-1)^n (P_{\bar{\xi}}P_{\xi})^n \right)
    \nonumber\\
\,&=&\,(1-(-1)^n)\,Tr\left((P_{\xi}P_{\bar{\xi}})^n\right)
\end{eqnarray}
proving that
\begin{eqnarray}
Tr[P_{\xi},P_{\bar{\xi}}]^n = 0 \hspace{5mm}\mbox{for}\hspace{5mm} n
                                \hspace{5mm}\mbox{odd},
\end{eqnarray}
and thus that ${\cal M}$ in \Ref{minpars} is identically zero.

As a result, the energy density \Ref{energyans} simplifies further
and if we treat ${\cal N}$ and  ${\cal I}$ as two parameters
then one can minimise the energy $\tilde{E}$ by solving the following 
Euler-Lagrange equations for $g$
\begin{eqnarray}
\label{eqng}
g_{rr} \left( 1+2\,{\cal N}\,\frac{1-\lambda}{A_N} \,\frac{\sin^2g}{r^2}
          + \,{\cal I}\,\frac{\lambda}{A_N}\,\frac{\sin^4g }{r^4}
      \right)
  +\frac{2}{r}\,g_r\,\left( 1-\,{\cal I}\,\frac{\lambda}{A_N}\,
                \frac {\sin^4g }{r^4}\right ) &&\nonumber
\\ \hspace{30mm}+ \frac{1}{A_N}\,\frac{\sin2g }{r^2} 
   \left ( {\cal N}\,((1-\lambda))g_{r}^{2}-1)
           +{\cal I}\,\frac{\sin^2g}{r^2}(\lambda\,g_{r}^{2}-1+\lambda)
   \right)=0.
\end{eqnarray}

We see from our analysis that the harmonic maps for the extended Skyrme model
are the same one as the usual Skyrme model. The harmonic map ansatz 
predicts thus that the solutions of the usual and the extended Skyrme 
models have the same symmetries. This has been confirmed by the numerical
solutions. The only difference, for the ansatz, between the two 
models comes from the profile function. This is due to the presence of the 
extra terms appearing in \Ref{eqng}.

In Table 1 we list the minimum energy for the harmonic maps that we will use 
later together with the corresponding value for $\cal{I}$.

\begin{table}[htbp]
\begin{center}
\begin{tabular}{|c||c |c||c|c| }
 \hline & \multicolumn {2}{c||}{} &  \multicolumn {2}{c|}{} \\
  & \multicolumn {2}{c||}{$SU(2)$}  &  \multicolumn {2}{c|}{$SU(3)$}\\
\hline & & & & \\ $B$ & Harmonic Map \, $f(\xi)$ & $\cal I$  &
Harmonic Map \, $f(\xi)$ & $\cal I$ \\ \hline & & & & \\ 1 &
$\left( \xi,1 \right)^t $ & 1 & $ \left(\xi,1 \right)^t $ & 1
\\
2 &  $ \left(\xi^2,1 \right)^t $   & 5.81 &  $
\left(\xi^2,\sqrt{2}\,\xi,1 \right)^t $ & 4
\\
3 &  $  \left(\xi(\xi^2-\sqrt{3}i),\sqrt{3}i\,\xi^2-1 \right)^t $
& 13.58 & $ \left(\xi^3,1.576\xi,\sqrt{2}\,^{-1} \right)^t $ &
10.65
\\
4 &  $  \left(\xi^4+2\sqrt{3}i\xi^2+1  ,\xi^4-2\sqrt{3}i\xi^2+1
\right)^t $ & 20.65 & $ \left(\xi^4,2.7191\xi^2,1 \right)^t $ &
18.05
\\
5 &  $  \left( \xi(\xi^4+b\xi^2+a) ,a\xi^4-b\xi^2+1 \right)^t, $ &
37.75 & $ \left(\xi^5-2.7\xi,2\,\xi^4+1,9/2\,\xi^3 \right)^t $ &
27.26
\\
 &  $  a=3.07\, , \, b=3.94 $ & & &
\\
  \hline   & & & & \\
5* &  $  \left(\xi(\xi^4-5) ,-5\xi^4+1 \right)^t $ & 52.05 & &
\\
5** &  $  \left(\xi^5 ,1 \right)^t $ & 84.425 & &
\\
 \hline
\end{tabular}
\end{center}
\caption{ Harmonic maps $f(\xi)$ minimising the angular
integral ${\cal I}$  for $SU(2)$ \cite{Manton} and $SU(3)$ \cite{sun}.
The 5* and 5** configurations denote saddle points that we also consider.}
\end{table}

\subsection{Energy and radius ratios for the $SU(2)$ model}

In this section, we analyse how the properties of the 
multi-Skyrmions rational map ansatz depend on the parameter $\lambda$.  
Using the value of ${\cal I}$ given in Table 1, we compute the 
profile $g$ by solving \Ref{eqng} and evaluate both the total energy and the 
radius of the configurations. In Figures 6 to 9, we show the energy ratio and 
the radius ratio defined in \Ref{ratio} for different values of the baryon 
number. At this stage we would like to remind the reader that 
$\lambda = 0$ corresponds to the pure Skyrme model while 
$\lambda = 1$ is equivalent to the pure Sk6 model.
Moreover, the ratio presented on the figures only depends on $\lambda$
{\it i.e.} the mixing between the two Skyrme terms.
When $B\geq6$, the graphs we obtained were all similar to Figure 5.

When comparing these results with the numerical solutions,
we notice first of all that the energy ratio predicted by the 
ansatz is always too large. Apart from this, the prediction for the energy 
is rather good except for the case $B=2$ where the energy difference 
between the numerical solution and the rational map ansatz is 7 times as large
for the Sk6 model than for the pure Skyrme model.

We also notice that the graphics obtained for the numerical solutions do not 
exhibit local minima as observed on the graphs obtained for the harmonic map 
ansatz. The only exception is the radius ratio obtained for the $B=5$ 
exact solution but the effect is so small that it could be a numerical 
artefact. 

The radius ratio obtained with the harmonic map ansatz is always too large
when compared with the radius ratio of the exact solution. For $B=2$, the 
radius
ratio increases with $\lambda$  and the error only gets worse as $\lambda$
increases. The case $B=3$ is rather surprising as the radius ratio has a 
deep local minimum around the value $\lambda=0.3$; this is where the relative 
error is the smallest, otherwise the relative error is smaller for the pure 
Sk6 model than for the pure Skyrme model. The cases $B=4$ and $B=5$ are very 
similar: the radius ratios decrease when $\lambda$ increases and the error 
for the pure Sk6 model is very small especially when $B=4$. 

We can thus conclude that the harmonic map ansatz produces good approximations
to the solutions of the generalised Skyrme model and the error is in most cases
smaller for the pure Sk6 model than for the pure Skyrme model, the only 
exception being the case $B=2$.

\vskip5mm
\begin{figure}[htbp]
\unitlength1cm \hfil
\begin{picture}(16,6)
 \epsfxsize=8cm \epsffile{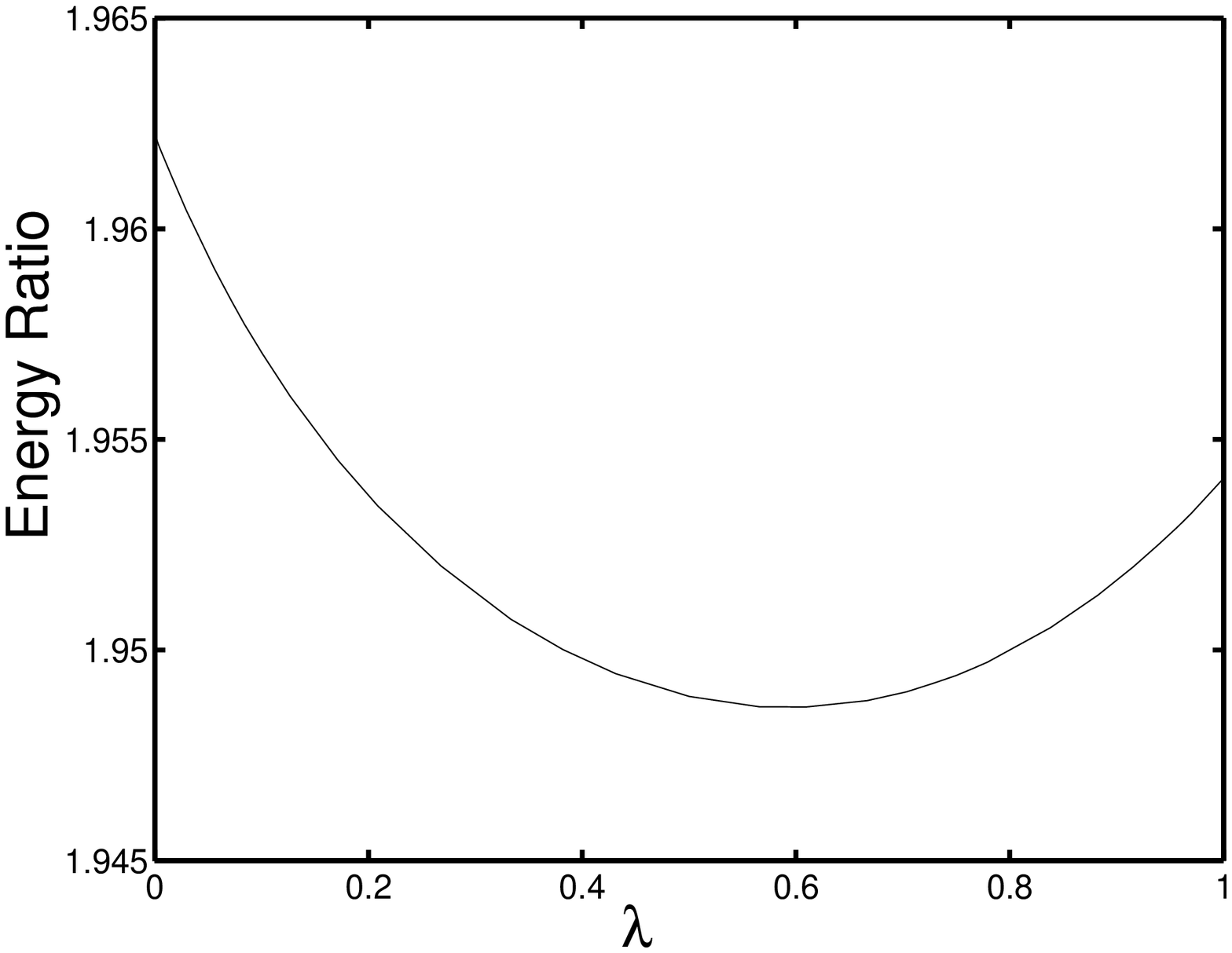}
 \epsfxsize=8cm \epsffile{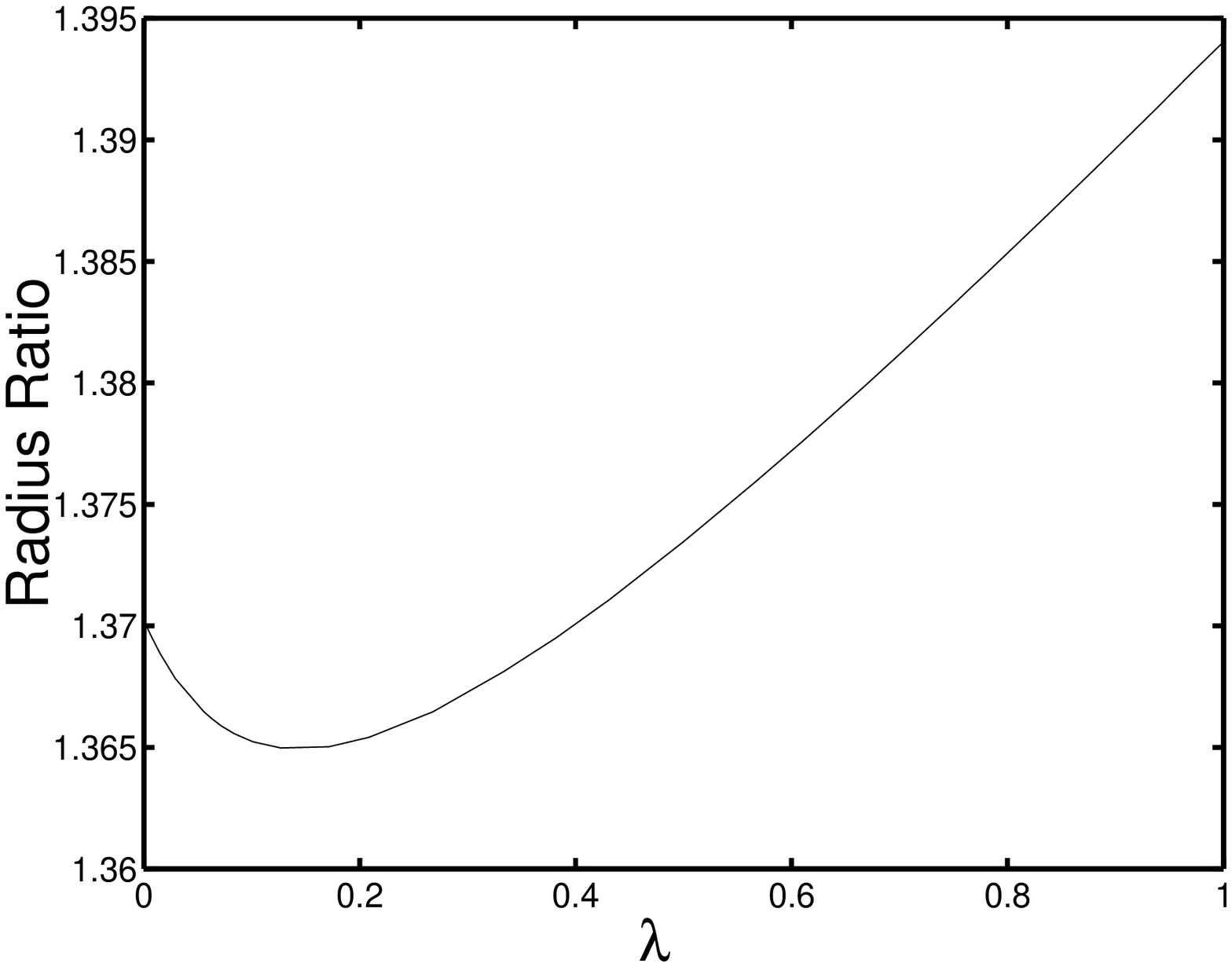}
\end{picture}
\caption{$\tilde{E}$ and $\tilde{R}$ ratio of
$\hspace{2mm}B=2/B=1\hspace{2mm}$ as a function of $\lambda$.}
\end{figure}

\vskip5mm
\begin{figure}[htbp]
\unitlength1cm \hfil
\begin{picture}(16,6)
 \epsfxsize=8cm \epsffile{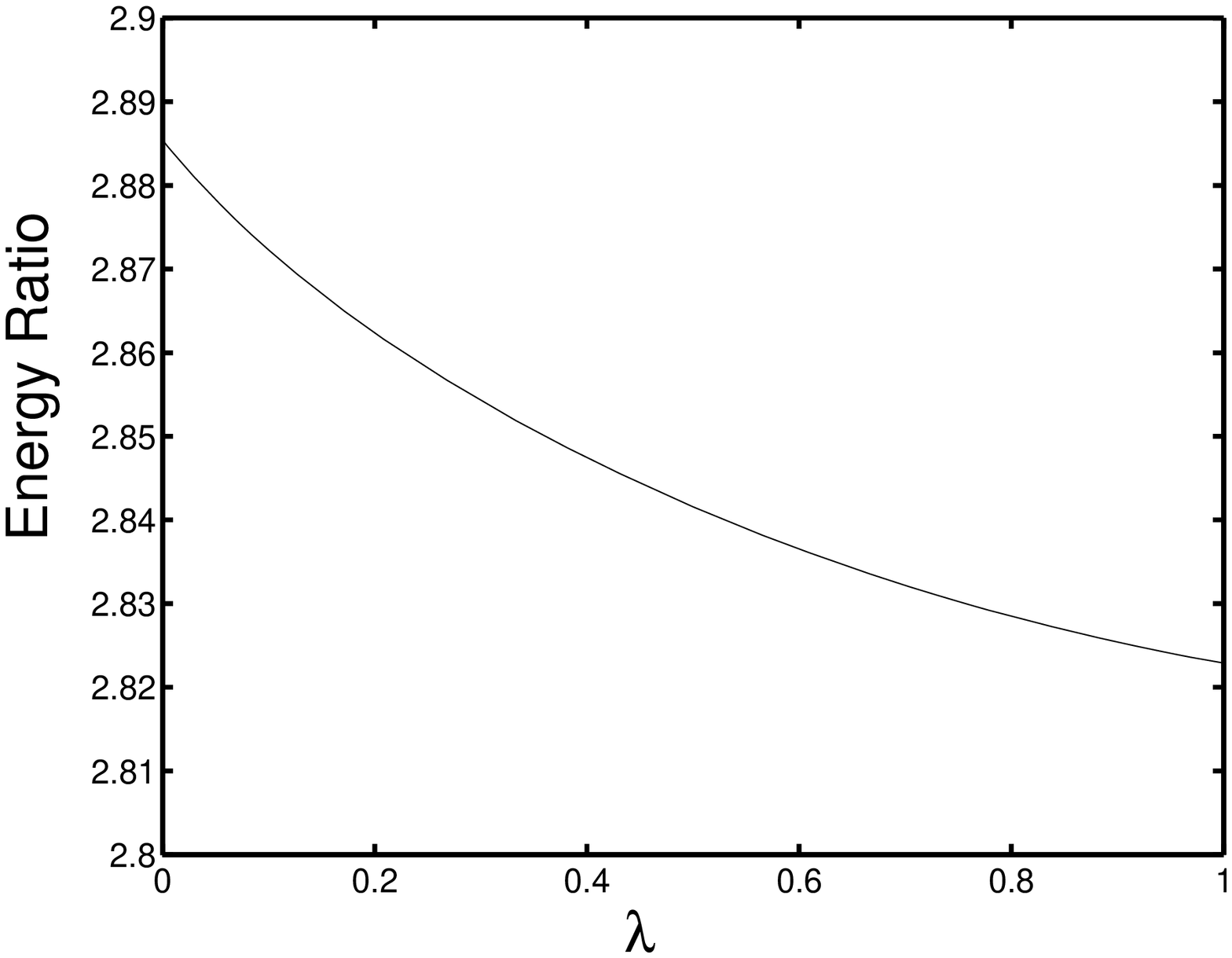}
 \epsfxsize=8cm \epsffile{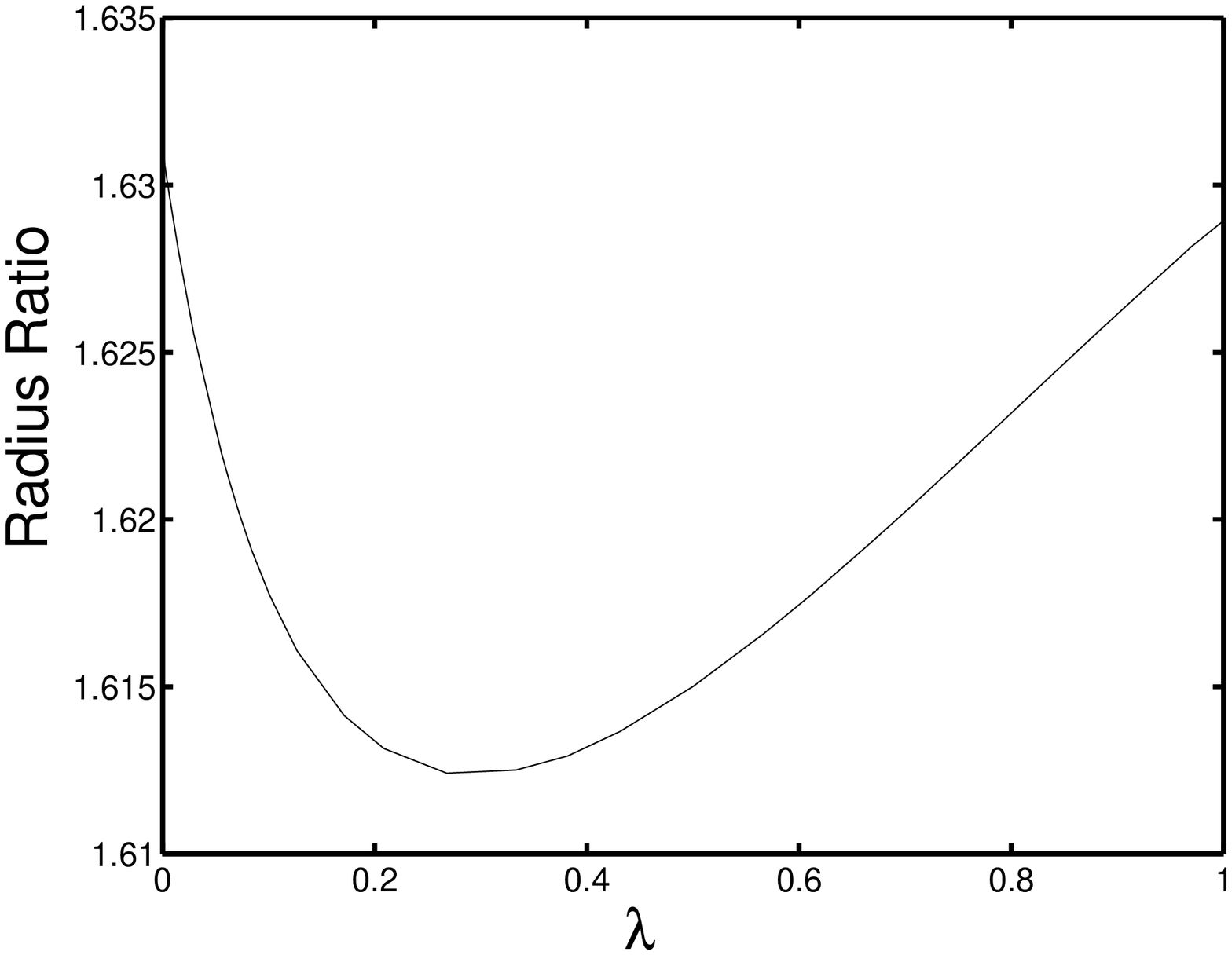}
\end{picture}
\caption{$\tilde{E}$ and $\tilde{R}$ ratio of
$\hspace{2mm}B=3/B=1\hspace{2mm}$ as a function of $\lambda$.}
\end{figure}

\vskip 5mm
\begin{figure}[htb]
\unitlength1cm \hfil
\begin{picture}(16,6)
 \epsfxsize=8cm \epsffile{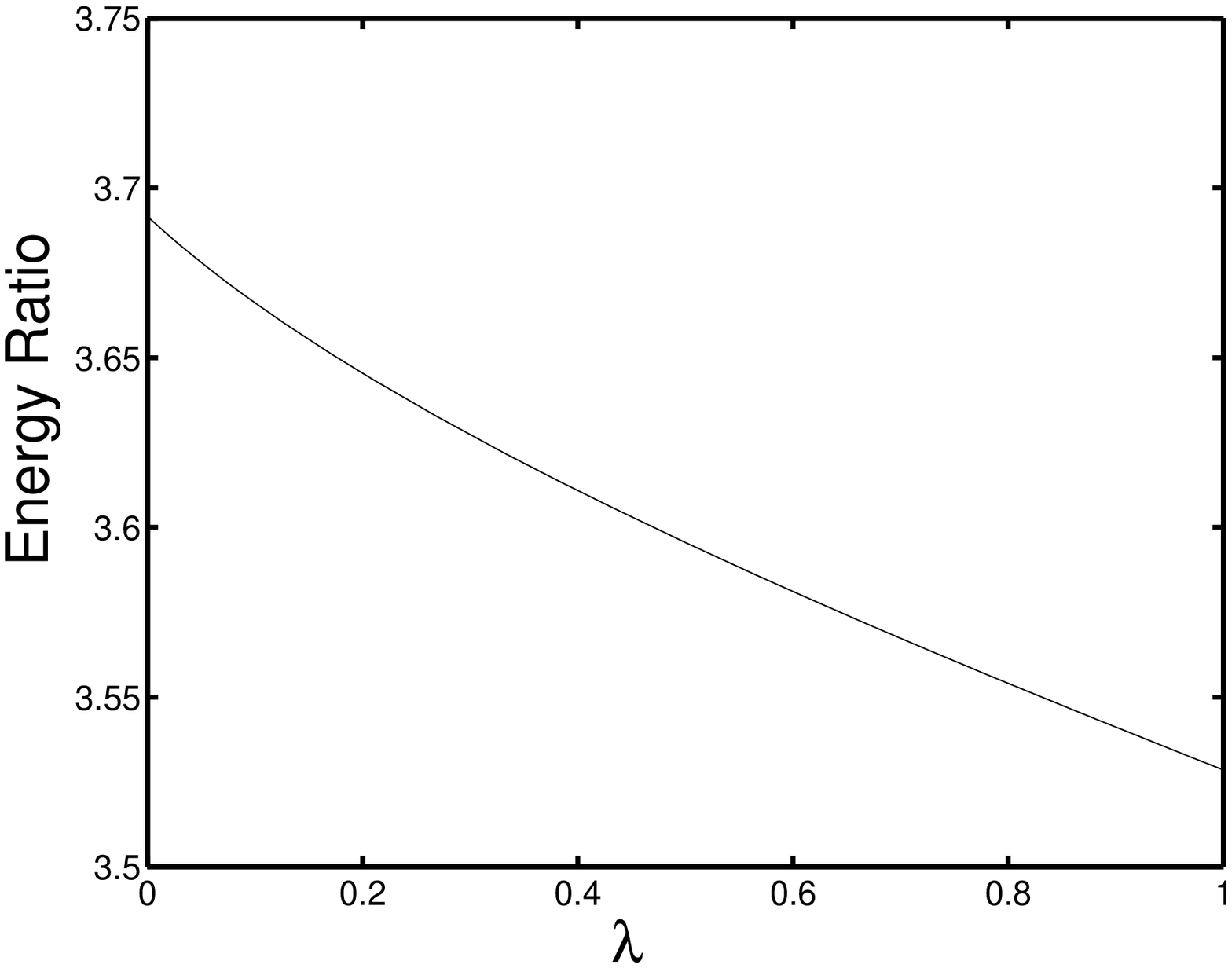}
 \epsfxsize=8cm \epsffile{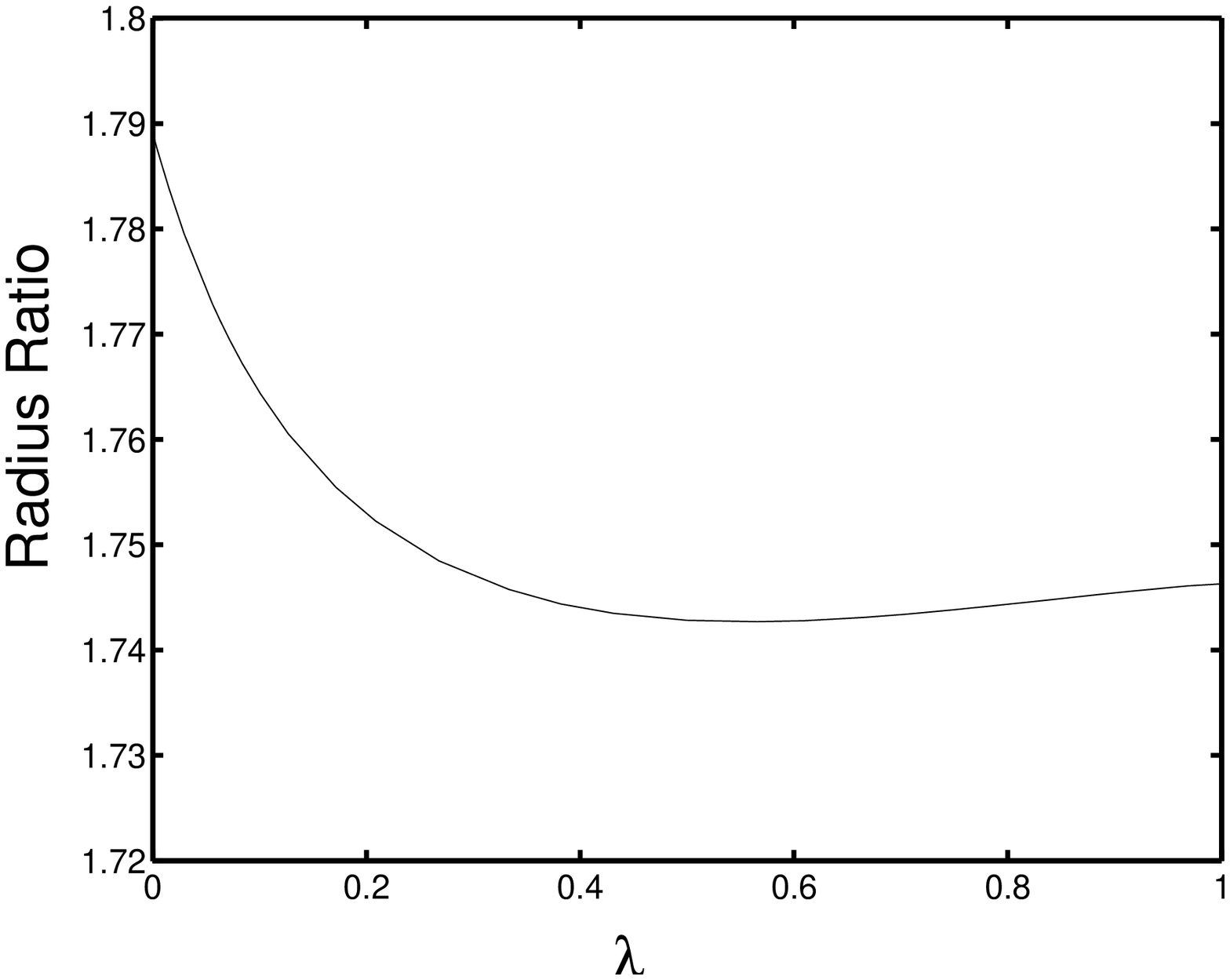}
\end{picture}
\caption{$\tilde{E}$ and $\tilde{R}$ ratio of
$\hspace{2mm}B=4/B=1\hspace{2mm}$ as a function of $\lambda$.}
\end{figure}

\vskip 5mm
\begin{figure}[htbp]
\unitlength1cm \hfil
\begin{picture}(16,6)
 \epsfxsize=8cm \epsffile{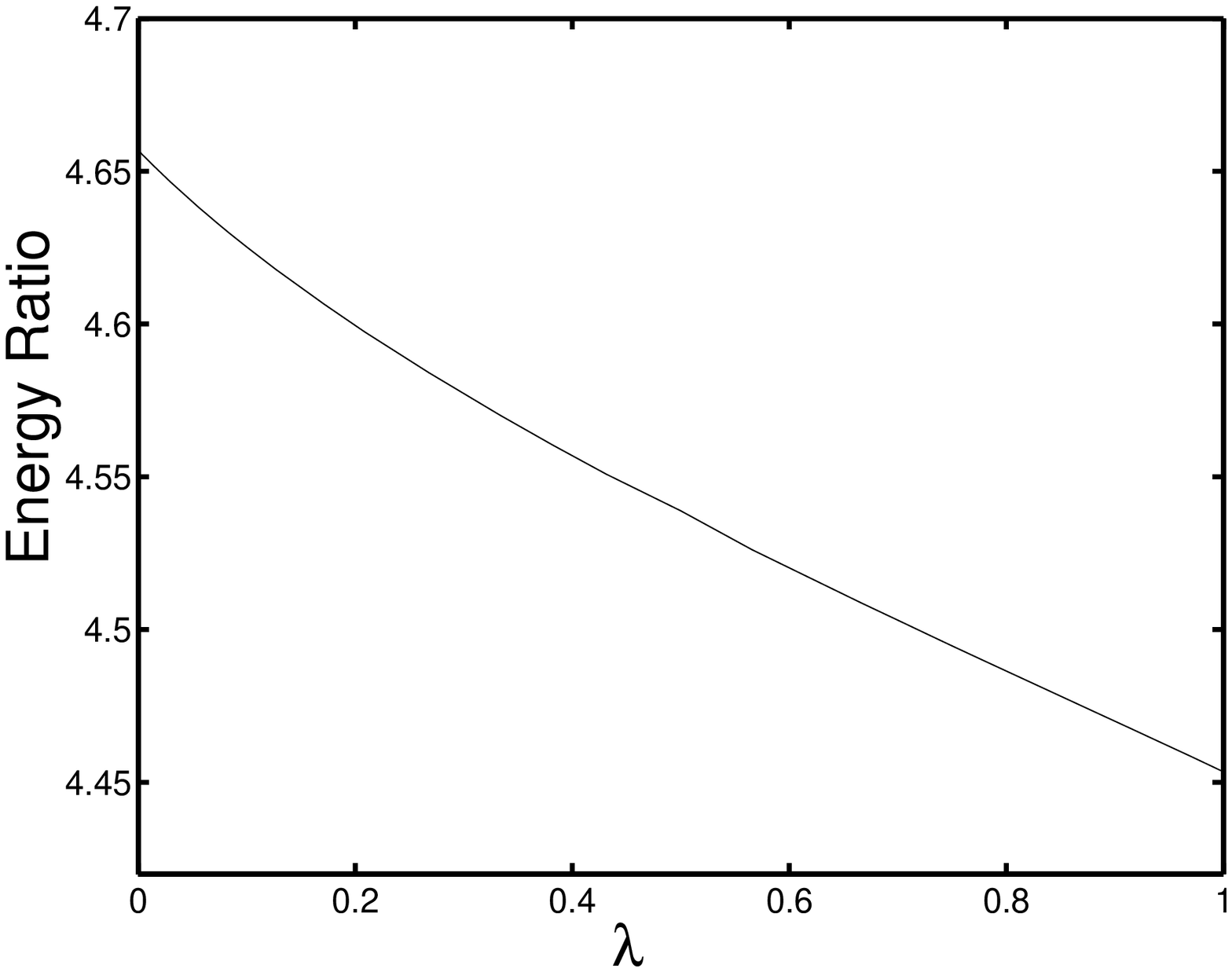}
 \epsfxsize=8cm \epsffile{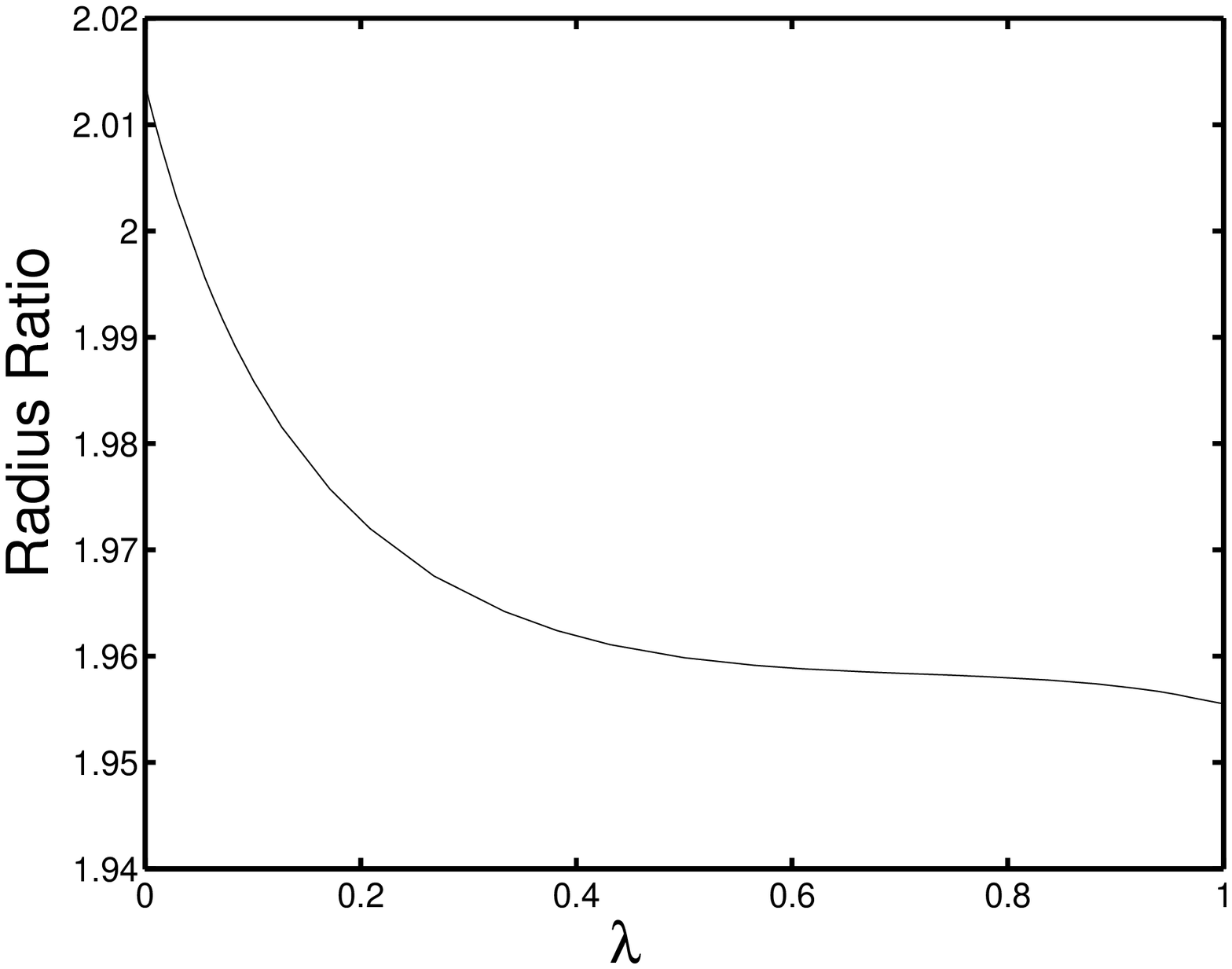}
\end{picture}
\caption{$\tilde{E}$ and $\tilde{R}$ ratio of
$\hspace{2mm}B=5/B=1\hspace{2mm}$ as a function of $\lambda$.}
\end{figure}

\vskip5mm
\begin{figure}[htbp]
\unitlength1cm \hfil
\begin{picture}(16,6)
 \epsfxsize=8cm \epsffile{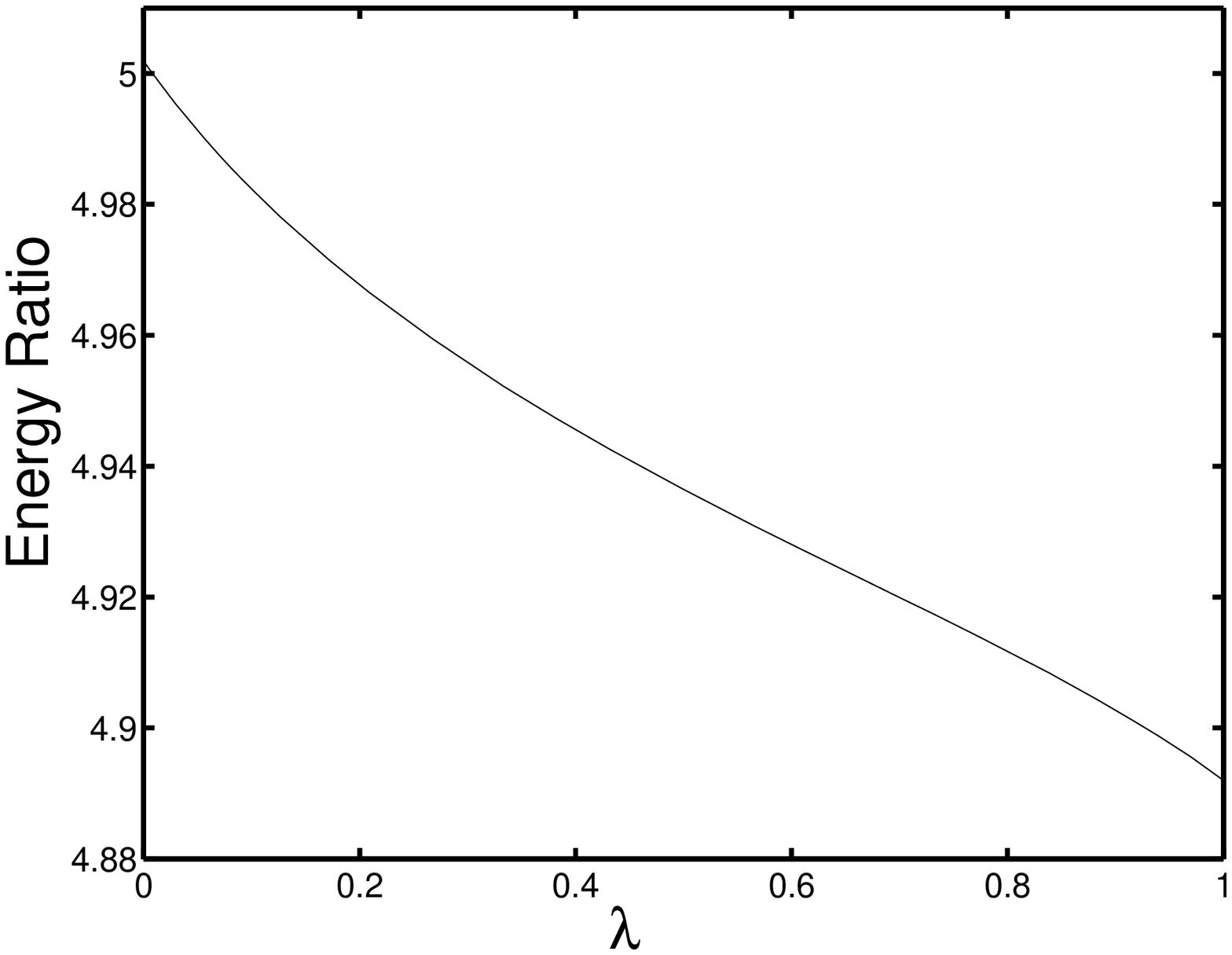}
 \epsfxsize=8cm \epsffile{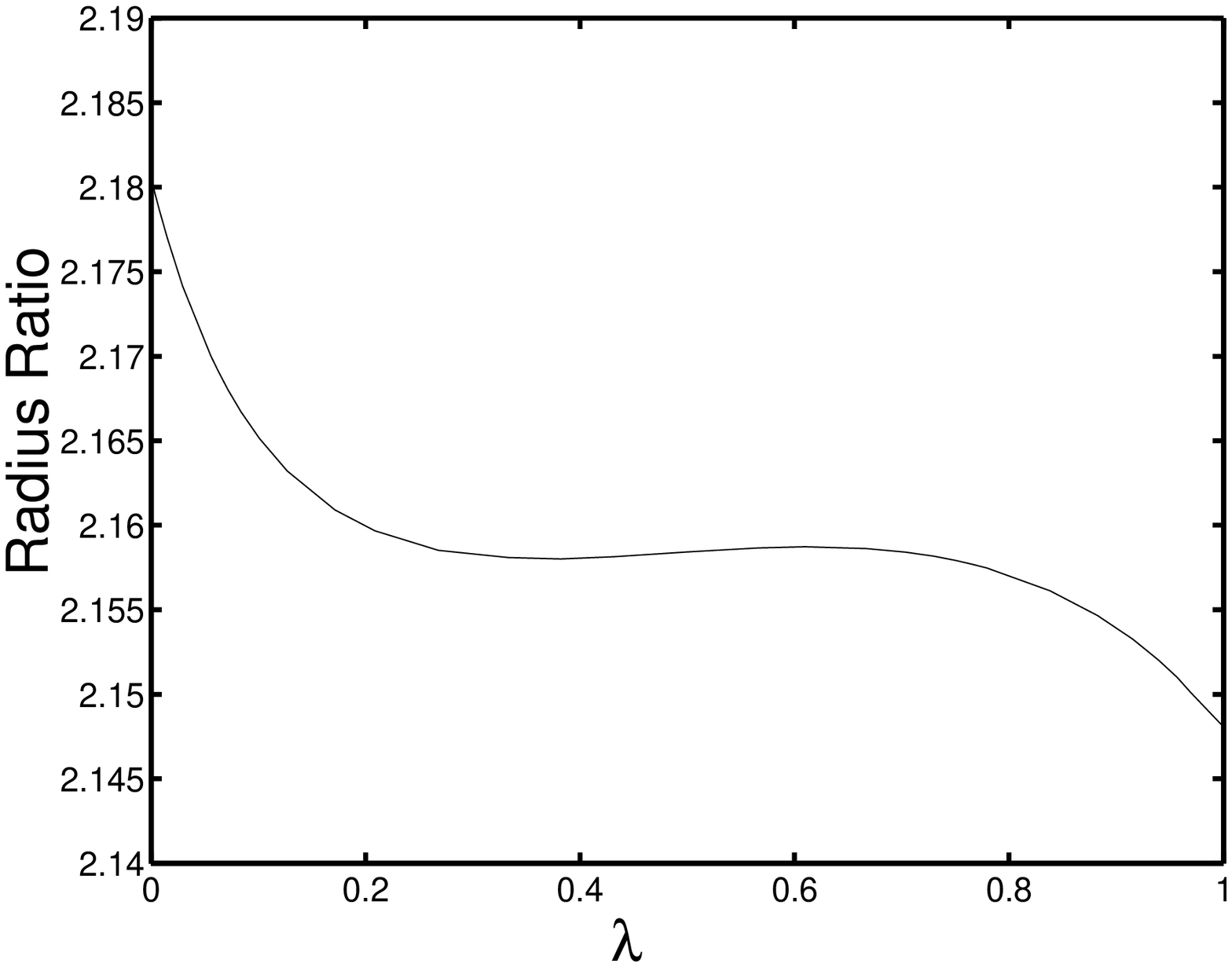}
\end{picture}
\caption{$\tilde{E}$ and $\tilde{R}$ ratio of
$\hspace{2mm}B=5^*/B=1\hspace{2mm}$ as a function of $\lambda$.}
\end{figure}

\vskip5mm
\begin{figure}[htbp]
\unitlength1cm \hfil
\begin{picture}(16,6)
 \epsfxsize=8cm \epsffile{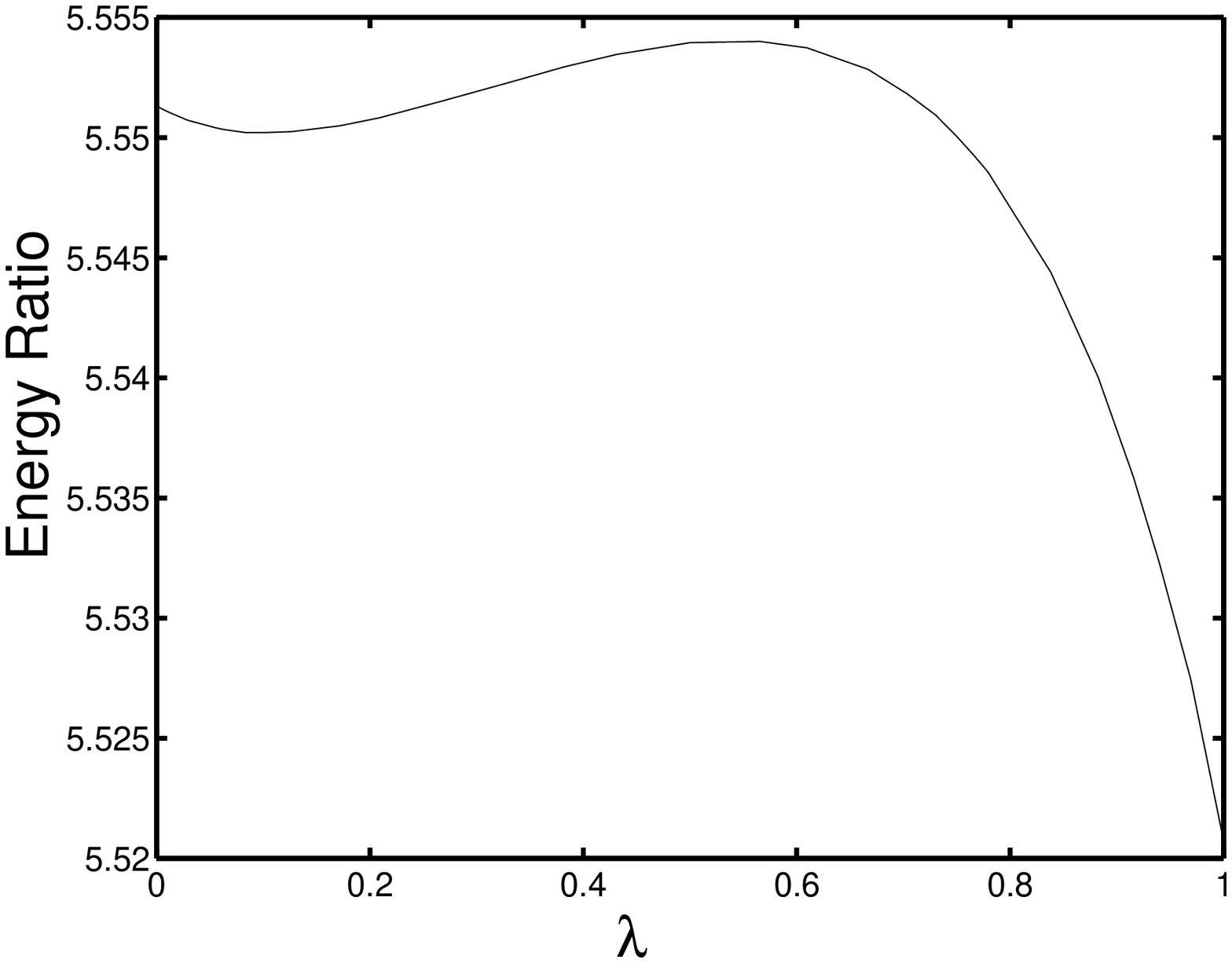}
 \epsfxsize=8cm \epsffile{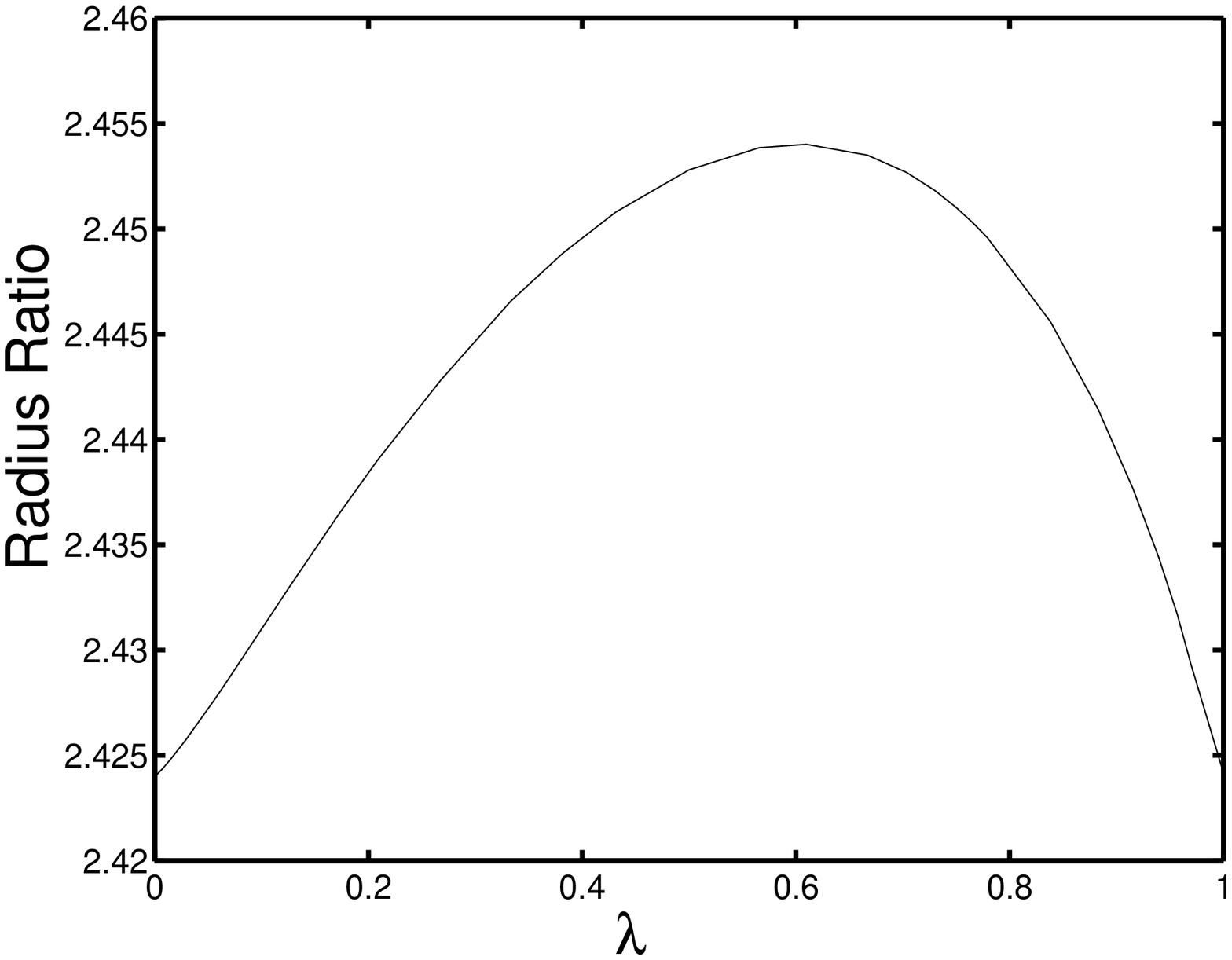}
\end{picture}
\caption{$\tilde{E}$ and $\tilde{R}$ ratio of
$\hspace{2mm}B=5^{**}/B=1\hspace{2mm}$ as a function of $\lambda$.}
\end{figure}

So far we have examined the behaviour of the model for harmonic maps
that minimise  the angular integral ${\cal I}$ and correspond to
minimum energy configurations. We have next considered harmonic maps that 
correspond to saddle points of the energy for $B=5$. The reason behind 
this selection lies on the fact that the 
binding energies of the multi-Skyrmion solutions are much larger than the
experimental values.
For this reason, we have considered the two harmonic maps, $B=5^*$  and 
$B=5^{**}$, given in Table 3. The first one has octahedral symmetry  
whereas the second gives a toroidal Skyrme field \cite{Manton}.

For  the case of $B=5^*$, shown on Figure 10,  we see that the binding energy 
is slightly larger than $5$ for the pure Skyrme model and that it decreases 
when the strength of the sixth-order term increases, going through the 
experimental value $4.97$ when $\lambda \approx 0.1$.

The second case, shown on Figure 11, is the only example where we have seen a 
local maxima for the energy ratio that is larger than the energy of both the
pure Skyrme and Sk6 model.

\subsection{Energy and radius ratios for the $SU(3)$ model.}

In this section we look at the harmonic maps configurations for the $SU(3)$
models\cite{sun}. The harmonic maps that we will use and the
corresponding values of ${\cal I}$ are all given in Table 3. The
single $SU(3)$ skyrmion is the well-known hedgehog ansatz and it is 
just an embedding of the $SU(2)$ solution.

Notice that the numerical constant $A_N$ appearing in \Ref{eqng} and 
\Ref{energyans} is now equal to $4/3$.

We should stress here that these configurations approximate solutions that
are believed to be saddle points of the energy. Their energy is larger than
the corresponding $SU(2)$ embeddings and they have a different symmetry as 
well.

\vskip5mm
\begin{figure}[htbp]
\unitlength1cm \hfil
\begin{picture}(16,6)
 \epsfxsize=8cm \epsffile{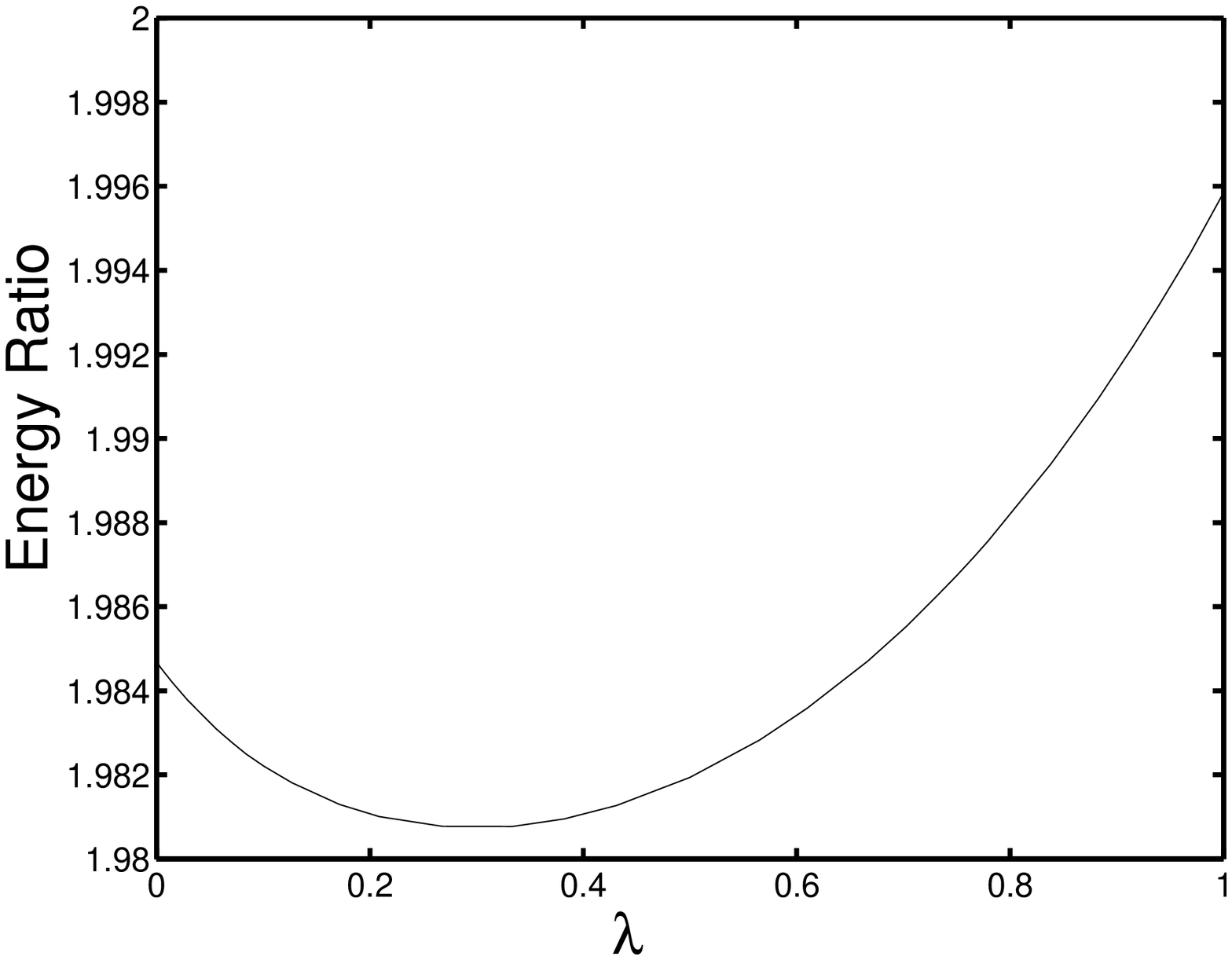}
 \epsfxsize=8cm \epsffile{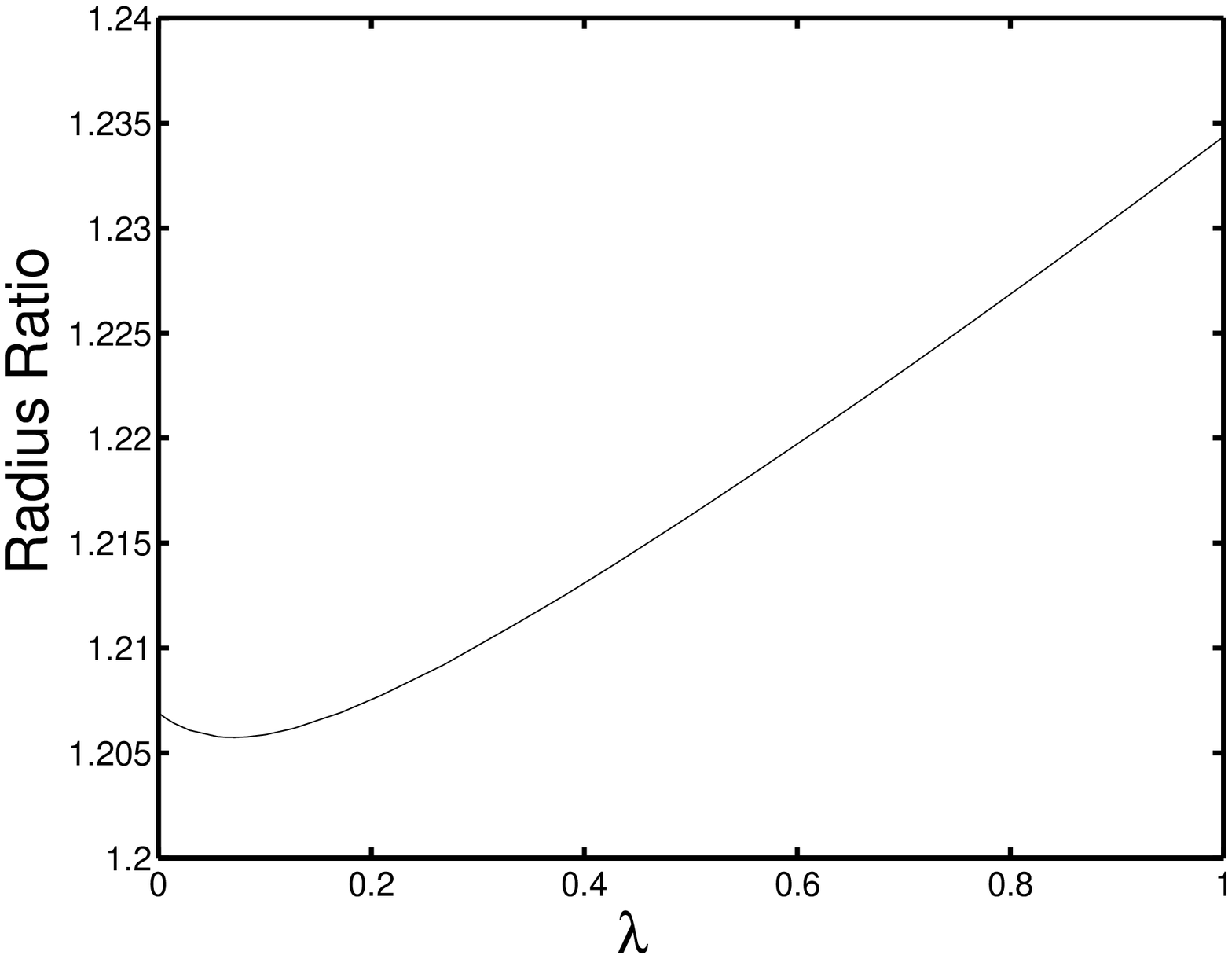}
\end{picture}
\caption{$\tilde{E}$ and $\tilde{R}$ ratio of
$\hspace{2mm}B=2/B=1\hspace{2mm}$ as a function of $\lambda$
for the $SU(3)$ harmonic map ansatz.}
\end{figure}

\vskip5mm
\begin{figure}[htbp]
\unitlength1cm \hfil
\begin{picture}(16,6)
 \epsfxsize=8cm \epsffile{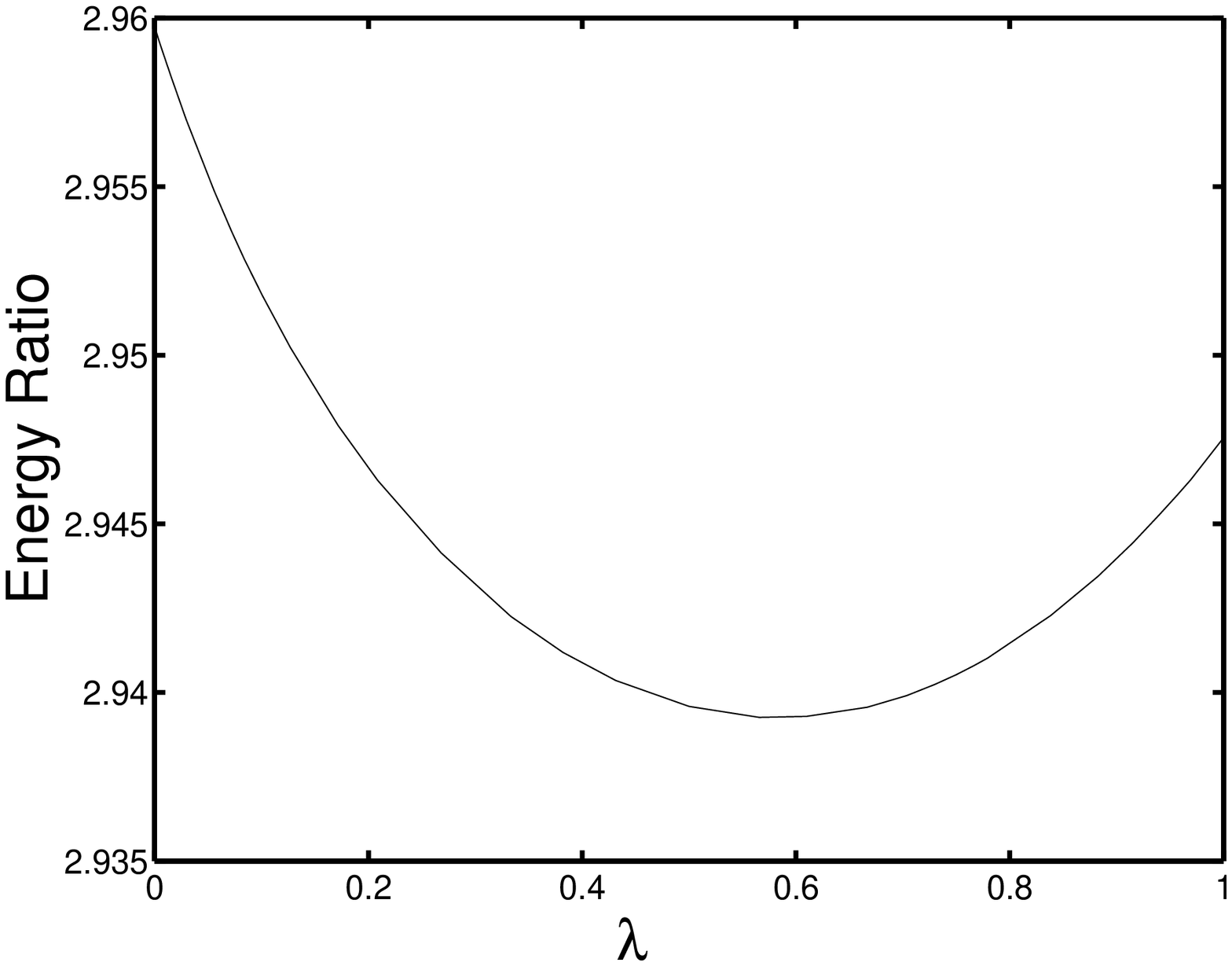}
 \epsfxsize=8cm \epsffile{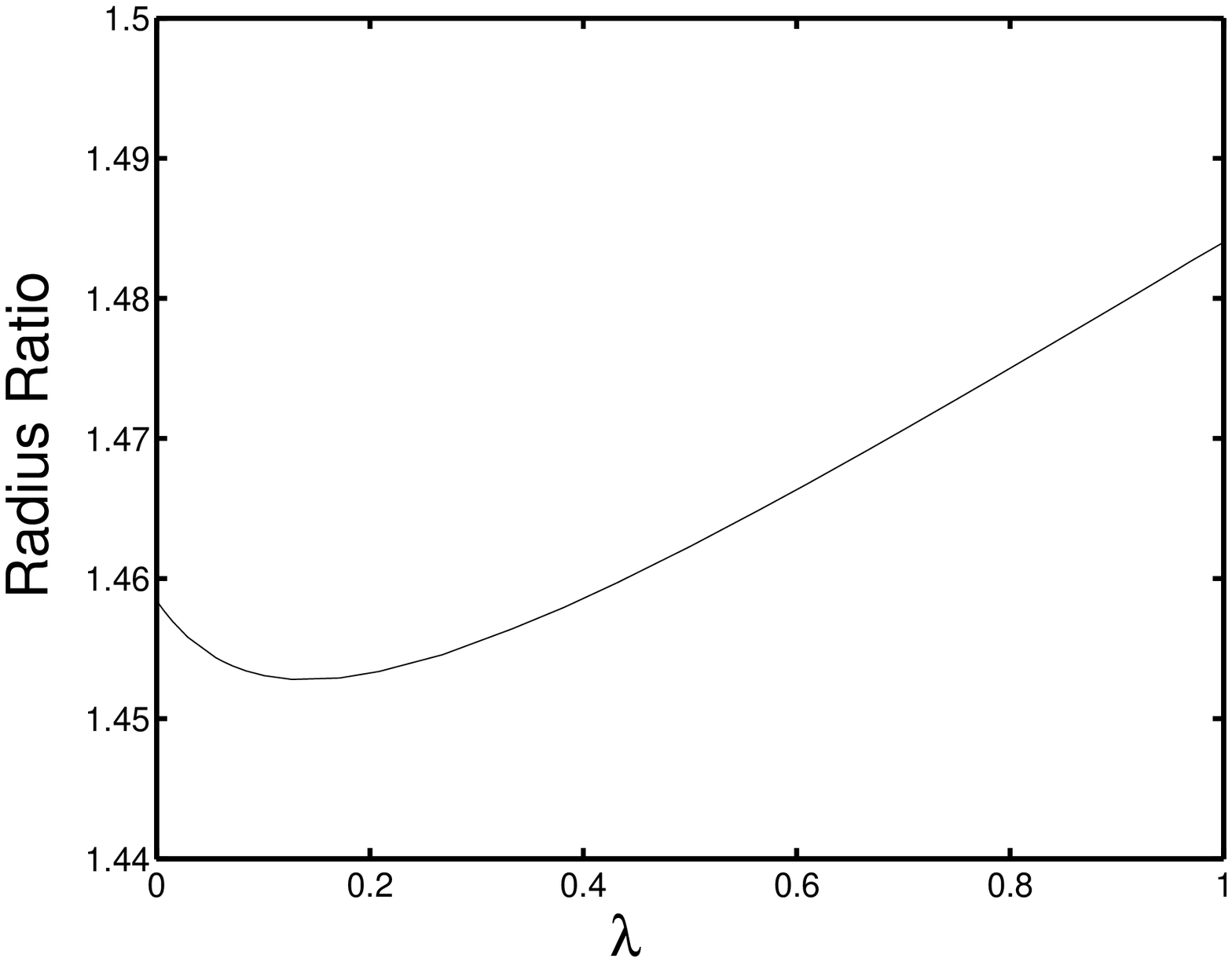}
\end{picture}
\caption{$\tilde{E}$ and $\tilde{R}$ ratio of
$\hspace{2mm}B=3/B=1\hspace{2mm}$ as a function of $\lambda$
for the $SU(3)$ harmonic map ansatz.}
\end{figure}

\vskip5mm
\begin{figure}[htbp]
\unitlength1cm \hfil
\begin{picture}(16,6)
 \epsfxsize=8cm \epsffile{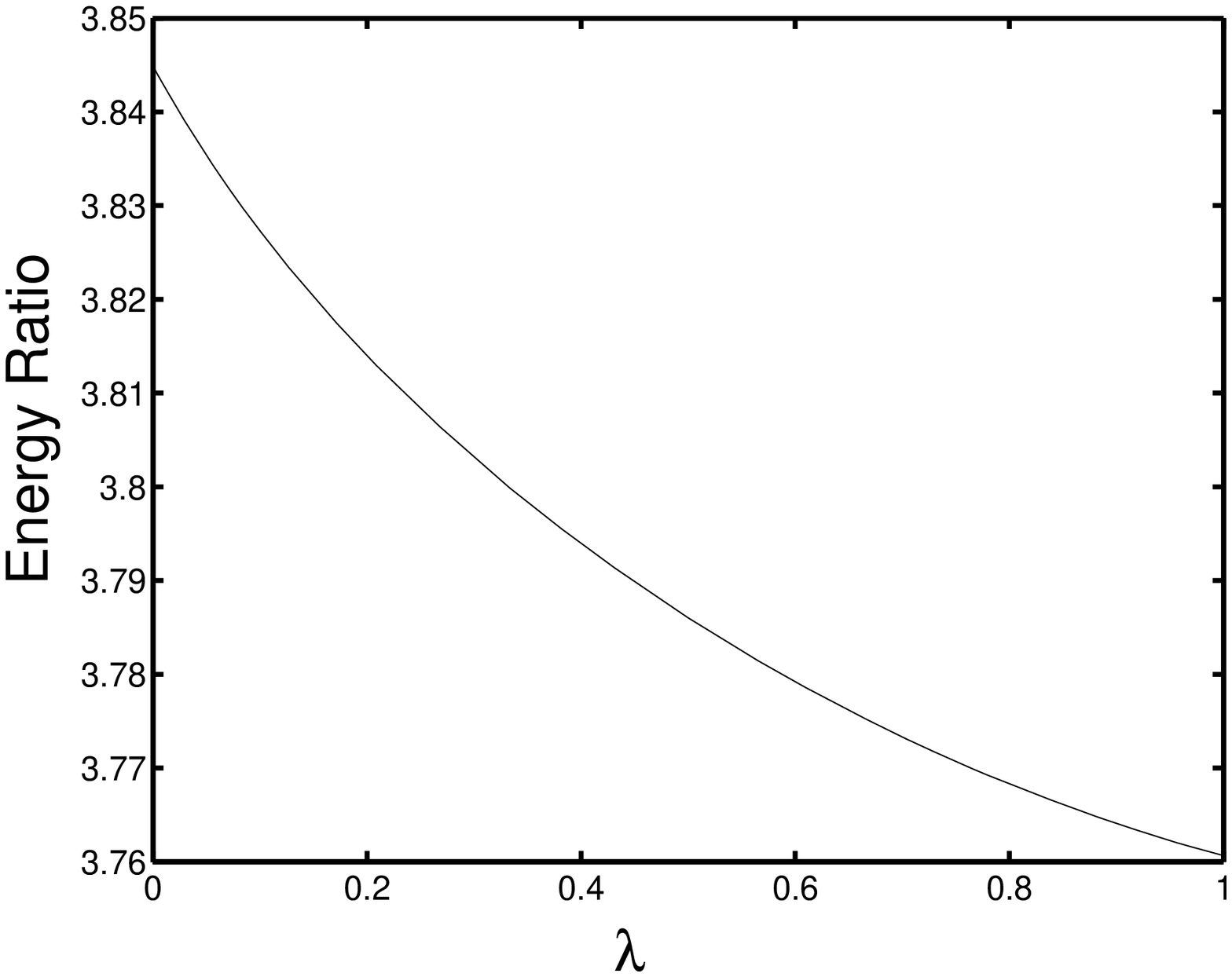}
 \epsfxsize=8cm \epsffile{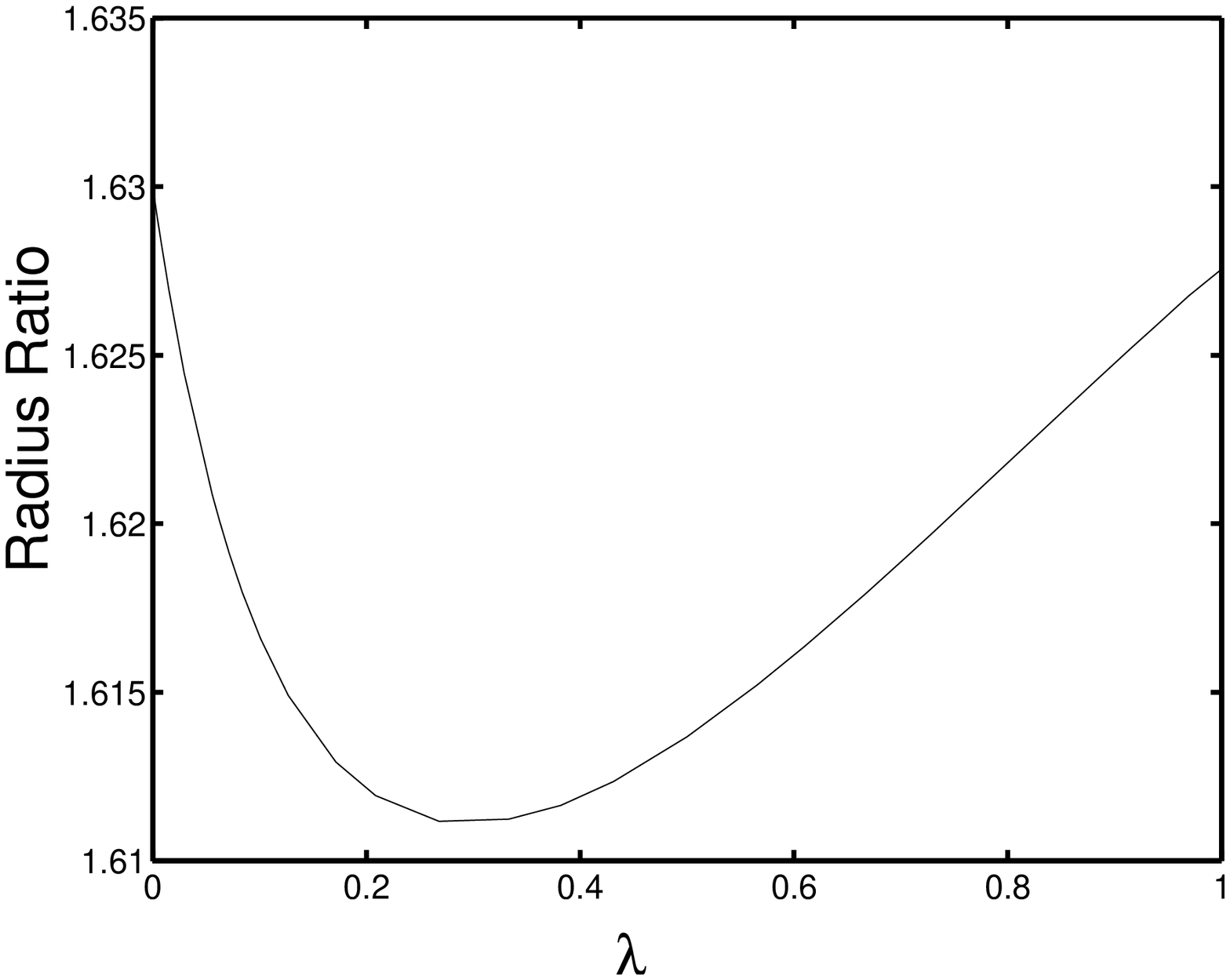}
\end{picture}
\caption{$\tilde{E}$ and $\tilde{R}$ ratio of
$\hspace{2mm}B=4/B=1\hspace{2mm}$ as a function of $\lambda$
for the $SU(3)$ harmonic map ansatz.}
\end{figure}

\vskip 5mm
\begin{figure}[htbp]
\unitlength1cm \hfil
\begin{picture}(16,6)
 \epsfxsize=8cm \epsffile{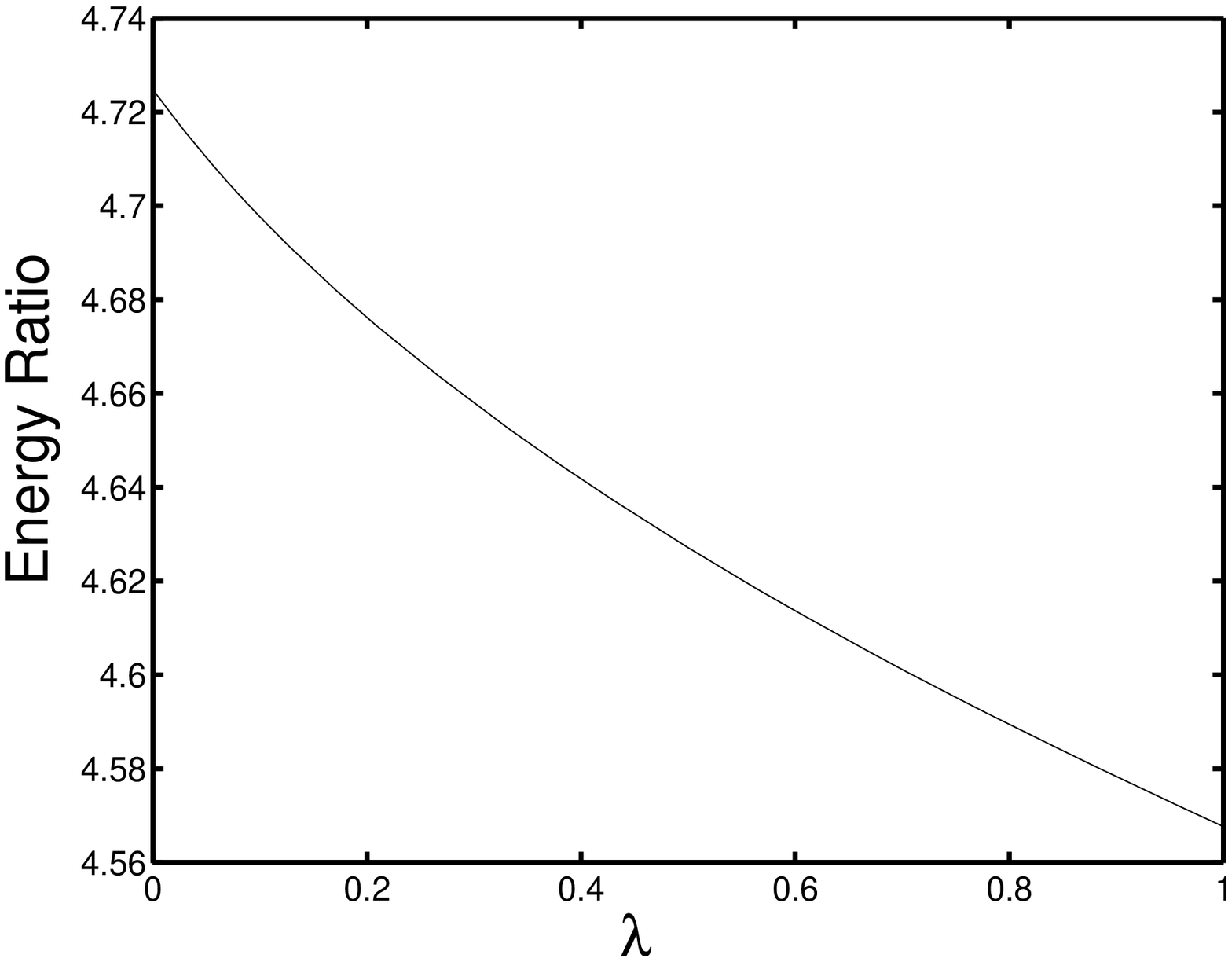}
 \epsfxsize=8cm \epsffile{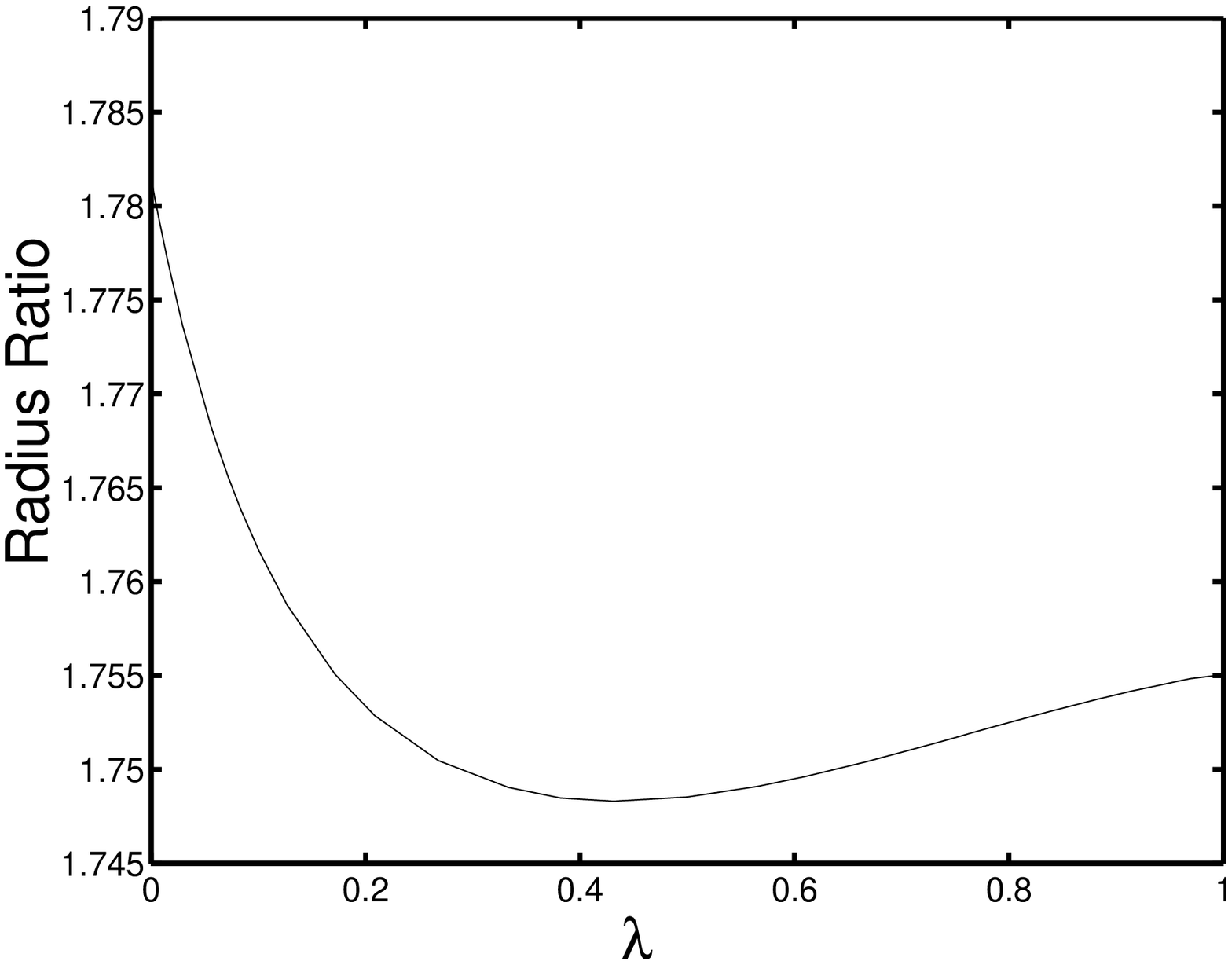}
\end{picture}
\caption{$\tilde{E}$ and $\tilde{R}$ ratio of
$\hspace{2mm}B=5/B=1\hspace{2mm}$ as a function of $\lambda$
for the $SU(3)$ harmonic map ansatz.}
\end{figure}

\vskip5mm
\begin{figure}[htbp]
\unitlength1cm \hfil
\begin{picture}(16,6)
 \epsfxsize=8cm \epsffile{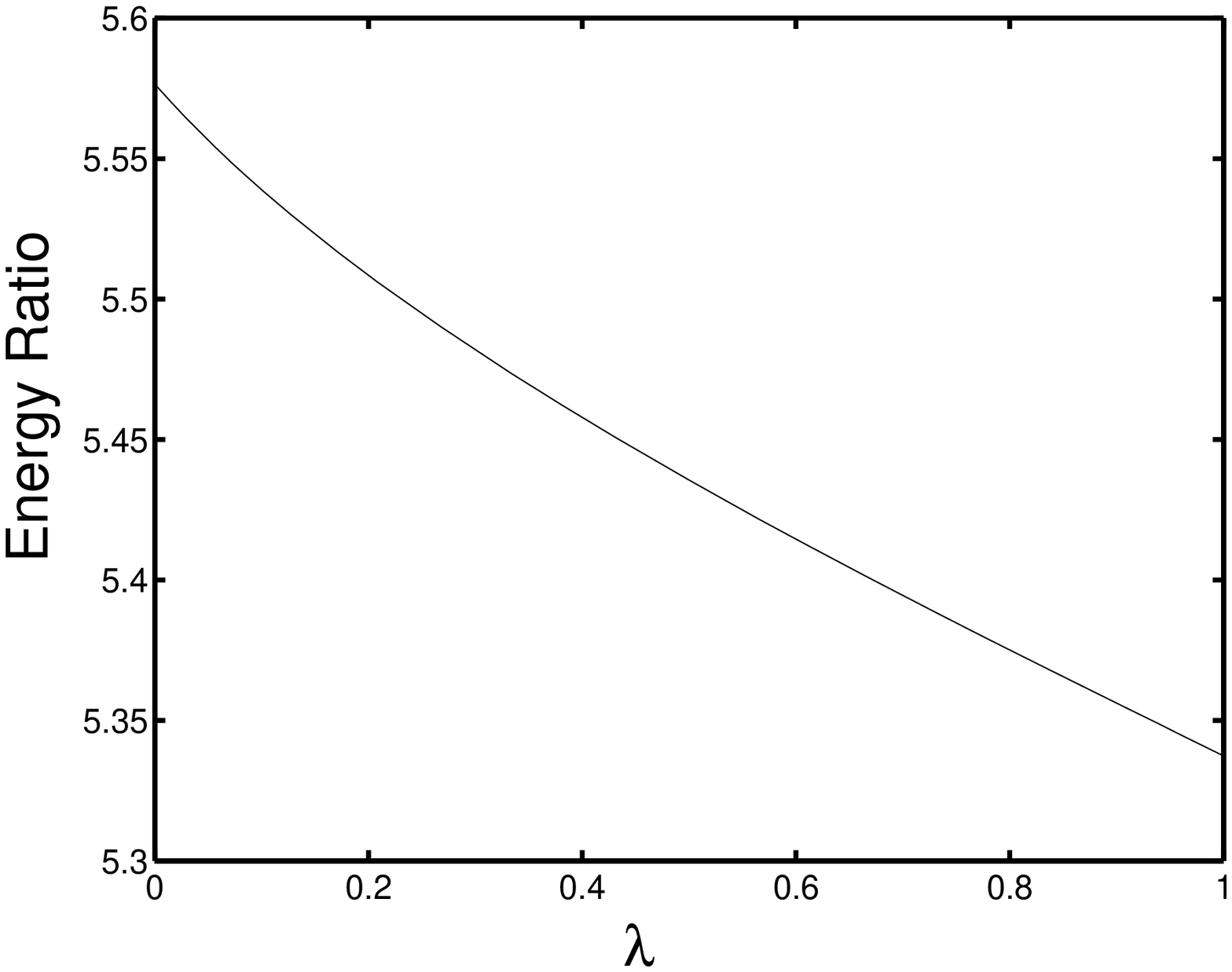}
 \epsfxsize=8cm \epsffile{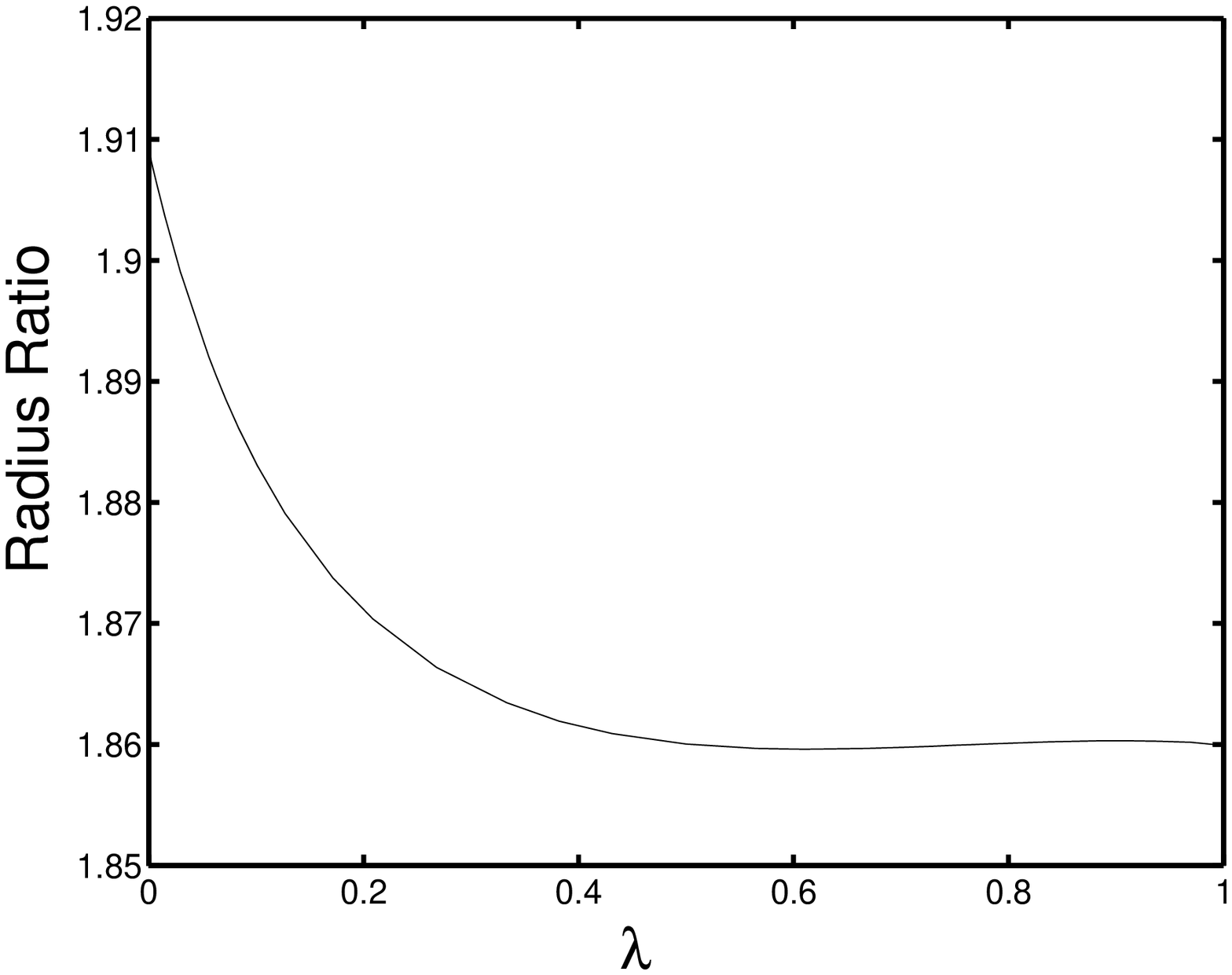}
\end{picture}
\caption{$\tilde{E}$ and $\tilde{R}$ ratio of
$\hspace{2mm}B=6/B=1\hspace{2mm}$ as a function of $\lambda$
for the $SU(3)$ harmonic map ansatz.}
\end{figure}

It is interesting to notice that unlike the $SU(2)$ model, the energy of the 
$B=2$ solutions increases with $\lambda$. 
For a given $B$ and a fixed value of $\lambda$, the energy ratio of these 
configurations is always larger than the energy ratio of the corresponding
$SU(2)$ solutions, while on the other hand, the radius ratios is always
smaller. 

It is also interesting to notice that the $\lambda$ dependence of the energy 
and radius ratios obtained for a given $B$ looks very much like the curve 
obtained for the $SU(2)$ model for $B-1$ Skyrmions. This can be explained 
by performing the change of variable $r \rightarrow r k$, where 
$k^2 = \frac{(1-\lambda)}{2}(1+\sqrt{1+ 4 \lambda /(A_N (1-\lambda)^2)})$,
and rewrite \Ref{energyans} as
\begin{eqnarray}
\label{modenergyans}
\tilde{E}={A_N k \over 3 \pi}  \int dr \hspace{3mm} 
 (A_N\, g_r^2 \,r^2 + 2{\cal N}'\, \sin^2g\, (1+(1-\lambda')g_r^2) 
       +(1-\lambda') {\cal I}'\, \frac{\sin^4g}{k^2 r^2} 
+\lambda'\,{\cal I}'\,\frac{\sin^4g}{r^2}\,g_r^2)&&\nonumber\\
&&
\end{eqnarray}
where ${\cal N}' = {\cal N} /A_N$, ${\cal I}' = {\cal I}/A_N$, and 
$\lambda' = \lambda/k^4$. The function $k(\lambda)$ monotonically
decreases from $k(0) = 1$ to $k(1) = A_N^{-1/4}$ and so it is relatively close 
to $1$ for all values of $\lambda$. In Table 4, we give the values
of  ${\cal N}'$ and ${\cal I}'$ for the $SU(3)$ ansatz and we notice that 
the $SU(3)$ solutions for $B=4$ and $B=5$ are closely related to the
$SU(2)$ solutions for respectively $B=3$ and $B=4$. 

\begin{table}[ht]
\begin{center}
\begin{tabular}{| c|c || c|c || c|c | }
\hline
\multicolumn{2}{|c||}{$SU(2)$} & \multicolumn{2}{c||}{$SU(3)$} 
&\multicolumn{2}{c|}{}  \\
\hline& & & & & \\ 
$N$ & $\cal{I}$& $N$ & $\cal{I}$&$\cal{N}'$ & $\cal{I}'$\\
\hline& & & & & \\ 
2 & 5.81  & 3 & 10.65 & 2.25 & 7.98\\ 
3 & 13.58 & 4 & 18.05 & 3    & 13.54\\
4 & 20.65 & 5 & 27.26 & 3.75 & 20.44\\
5 & 37.75 & 6 & 37.33 & 4.5  & 28 \\
\hline
\end{tabular}
\end{center}
\caption{${\cal N}'$ and ${\cal I}'$ for the $SU(3)$ anstaz.}
\end{table}

\begin{table}[ht]
\begin{center}
\begin{tabular}{|c||c |c||c|c||c|c| }
 \hline
 & \multicolumn{2}{c||}{} & \multicolumn{2}{c||}{} & \multicolumn{2}{c|}{}
\\& \multicolumn{2}{c||}{$SU(2)$ Numerical Solutions} & \multicolumn{2}{c||}
{$SU(2)$}& \multicolumn{2}{c|}{$SU(3)$}
\\ \hline & & & & & & \\ $B$ & Skyrme Ratio & Sk6 Ratio & Skyrme Ratio &
Sk6 Ratio & Skyrme Ratio & Sk6 Ratio
\\\hline & & & & & & \\

2 & 1.9009 & 1.8395 & 1.96223 & 1.95407 & 1.98468 & 1.99585\\

3 & 2.7650 & 2.7103 & 2.88541 & 2.82888 & 2.95974 & 2.94756\\

4 & 3.6090 & 3.5045 & 3.69164 & 3.52850 & 3.84491 & 3.76064\\

5 & 4.5000 & 4.3780 & 4.65685 & 4.45345 & 4.72485 & 4.56766\\

6 &    -   &  -     & 5.54105 & 5.26743 & 5.57660 & 5.33739\\

\hline
\end{tabular}
\end{center}
\caption{Energy ratio, $E_B/E_{B=1}$, for the $SU(2)$ numerical solutions,
and the $SU(2)$ and $SU(3)$ rational map anstaz configuration.}
\end{table}

\begin{table}[htbp]
\begin{center}
\begin{tabular}{|c||c |c||c|c||c|c| }
 \hline
 & \multicolumn{2}{c||}{} & \multicolumn{2}{c||}{} & \multicolumn{2}{c|}{}
\\& \multicolumn{2}{c||}{$SU(2)$ Numerical Solutions} & \multicolumn{2}{c||}
{$SU(2)$}& \multicolumn{2}{c|}{$SU(3)$}
\\ \hline & & & & & & \\ $B$ & Skyrme Ratio & Sk6 Ratio & Skyrme Ratio &
Sk6 Ratio & Skyrme Ratio & Sk6 Ratio
\\\hline & & & & & & \\

2 & 1.3549 & 1.308 & 1.37023 & 1.39403& 1.20691 & 1.234384\\

3 & 1.5080& 1.5570 & 1.63107 & 1.62894& 1.45842& 1.483996\\

4 & 1.6850 &1.7420 & 1.78911 &1.746286 & 1.63002&1.62755 \\

5 & 1.8890& 1.9250 & 2.013822& 1.95551& 1.78149 &  1.75505\\

6 & - & - & 2.178298&2.09768 &1.909141 &1.859916 \\

\hline
\end{tabular}
\end{center}
\caption{Radius ratio, $E_B/E_{B=1}$, for the $SU(2)$ numerical solutions,
and the $SU(2)$ and $SU(3)$ rational map anstaz configuration.}
\end{table}

In tables 5 and 6 we compare the energies and the radius ratios, obtained 
for the $SU(2)$ and $SU(3)$ model using the rational map ansatz.
We also compare these values with the $SU(2)$ numerical solutions.

\subsection{Conclusions}

We have studied an extension of the Skyrme model defined by adding to the
Lagrangian a sixth-order term. We have computed the multi-Skyrmion solutions
of the extended model for up to $B=5$ and we have shown that they have the 
same symmetry as the pure Skyrme model. 
We have analysed the dependence of the energy and radius
of the classical solution with respect to the coupling constant $\lambda$.
We found that the addition of the sixth-order term makes the multi-Skyrmion 
solution more bound than in the pure Skyrme model and that it also reduces the 
solution radius.

We have also used the harmonic map ansatz to approximate the numerical 
solutions and we found that the ansatz works as well, and in many cases even 
better, for the extended model than for the pure Skyrme model.
 
\appendix
\section{APPENDIX}
In this appendix we describe the numerical methods that we have used 
to compute our numerical solutions. 

\subsection{3 Dimensional Solutions}
To compute the three dimensional solutions described in section 2, we  
discretised the static equations using finite differences and we solved 
them using the relaxation method. We used the fixed boundary condition, 
taking the vacuum value for the field on the edge of the grid.

The values obtained for the energy and the radius with our methods are 
affected by  two sources of inaccuracy. The first one is the finiteness of 
the grid which, by distorting the field slightly, increases the value of the 
energy. The second one is the fact that finite differences systematically
underestimate the value of the energy. One could of course hope that the two
effects cancel out, but as it is difficult to evaluate their order of magnitude
one has to experiment and reduce them both as much as possible.

To reduce the edge effects, we computed the same solutions on grids of 
different
sizes $L$ but keeping the lattice spacing $dx = L/N $ constant, where $N$ is 
the number of lattice points in each direction. We then looked at how the
energy changed as a function of the size and chose a value for $L$
for which the energy is only  slightly affected by the edge effects.  

The finite difference scheme we used is of order two, so when we evaluate the 
total energy we can write
\begin{equation}
\label{ennum}
E = E_0 + E_1 dx + E_2 dx^2 + O(dx^3).
\end{equation}
In theory, $E_1 = 0$ but in practice it is a small but non-zero coefficient
induced by the edge effects. To improve the evaluation of the energy for a 
given solution, we computed the solution for at least three different values 
of $dx$ using a grid of size $L$ for which the edge effects are sufficiently 
small. We then fitted these values to the coefficient $E_0$, $E_1$ and $E_2$ 
in \Ref{ennum} getting $E_0$ as a better estimation for the energy. 
Notice also that $E_2$ is always negative and that 
$\mod{E1}*dx \ll \mod{E2}$. When this last condition is not 
satisfied one must conclude that the edge effects are large and one must 
increase $L$. To check our evaluation we performed the same interpolation
for the topological charge $Q = Q_0 + Q_1 dx + Q_2 dx^2 + O(dx^3)$. As we know
that it must be an integer $B$, the quantity $(B-Q_0)/B$
is a good estimation of the relative error on the topological charge but also
on the energy $E_0$. 

For the solution $B=2 \dots 5$ we used a box ranging from $-8$ to $8$ in all 
directions and we used grids of $100$ and $120$ and $140$ points. We also found
that for a given value of $B$, $E_2$ did not change much with $\lambda$. We 
were thus able to evaluate it for a few values of $\lambda$ and used an 
extrapolation for the other values. We also found that the 
relative error on the energy was smaller than $0.5\%$. 

As an alternative method to evaluate the energy we have considered computing
the quantity $E/Q$, as if $E_2$ and $Q_2$ were comparable, we would 
not have to compute the solutions for different values of $dx$. Unfortunately 
we found that with our discretisation, $Q_2$ is about $50\%$ larger than 
$E_2$ and as a result the value we get for $E/Q$ increases when $dx$ 
decreases and thus underestimates the energy value.

To evaluate the radius, we have used the same method, but for this quantity 
the integrand decreases more slowly towards infinity and as a result the
value is more affected by the finiteness of the grid. We believe that the
overall behaviour of the radius ratio graph can be trusted but some of 
the fine details might be numerical artefacts.   

To double-check our results, we have computed the $B=2$ 
axially symmetric solutions by solving \Ref{equationaf} and \Ref{equationag}
on a two-dimensional grid. This made it possible to use many more points and
much larger grids. When using the grid defined by $z \in [-20,20]$ and 
$r \in [0,20]$ taking $dx = 0.05$ the error was smaller than $0.1\%$ and we 
found for example $E=2.378$ for the pure Skyrme model. When computing this 
solution by solving the three-dimensional equation using the method described 
above, the difference between the two energies was less than $0.1\%$, 
thus validating the methods that we used. 

The energy values that we obtained for the pure Skyrme model all fit 
within $1\%$ the value given in \cite{BS} except for $B=2$ where the error
is $1.5\%$. As the numerical methods and the type of grid used are not
described in \cite{BS} it is difficult to make any further comparison between
the numerical results.    

\subsection{profiles}
To compute the profile functions for the hedgehog ansatz or the harmonic map
ansatz we have used both the shooting and the relaxation methods.
We have in every case compared the solutions obtained with grids of different 
sizes and different number of points to ensure that our results were accurate 
and that they were not affected by edge effects. We were led to use very 
large grids, $R_{max} =80$, to get an accurate value for the radius as well as
up to $160000$ lattice points.

\section{Acknowledgement}
One of us, I.F., would like to thank T. Weidig for useful discussions during 
the early part of this work. We would like to thank V. Piette for her 
linguistic advice.

\end{document}